\documentclass[journal=apchd5,manuscript=article]{achemso}

\usepackage[version=3]{mhchem} 
\usepackage{epsfig}
\usepackage{graphicx}
\usepackage{subfigure}
\usepackage{color}
\usepackage{soul}
\usepackage{float}
\usepackage{epstopdf}
\usepackage{amsmath}
\usepackage{amssymb}
\usepackage{ifthen}
\usepackage{alltt}
\usepackage{fancyhdr}
\usepackage{verbatim}
\usepackage{moreverb}
\usepackage{rotating}
\usepackage{colortbl}
\usepackage{amssymb}
\usepackage{multirow}
\usepackage{bigstrut}
\usepackage{wrapfig}

\def\e{\begin{equation}}
\def\f{\end{equation}}
\def\_#1{{\bf #1}}

\def\.{\cdot}

\def\=#1{\overline{\overline #1}}

\author{Y.~Ra'di}
 \email{younes.radi@aalto.fi}
 \affiliation{Department of Radio Science and Engineering, Aalto University, P.O. Box 13000, FI-00076 Aalto, Finland}
\author{V.~S.~Asadchy}
\affiliation{Department of Radio Science and Engineering, Aalto University, P.O. Box 13000, FI-00076 Aalto, Finland}
\author{S.~U.~Kosulnikov}
\affiliation{Department of Radio Science and Engineering, Aalto University, P.O. Box 13000, FI-00076 Aalto, Finland}
\author{M.~M.~Omelyanovich}
\affiliation{Department of Radio Science and Engineering, Aalto University, P.O. Box 13000, FI-00076 Aalto, Finland}
\author{D.~Morits}
\affiliation{Department of Radio Science and Engineering, Aalto University, P.O. Box 13000, FI-00076 Aalto, Finland}

\author{A.~V.~Osipov}
\affiliation{%
Microwaves and Radar Institute, German Aerospace Center (DLR), Germany}

\author{C.~R.~Simovski}
\affiliation{Department of Radio Science and Engineering, Aalto University, P.O. Box 13000, FI-00076 Aalto, Finland}

\author{S.~A.~Tretyakov}
\affiliation{Department of Radio Science and Engineering, Aalto University, P.O. Box 13000, FI-00076 Aalto, Finland}

\title[An \textsf{achemso} demo]
  {Full light absorption in single arrays of spherical nanoparticles}

\begin{document}

%
%
%
%
%

\begin{abstract}
In this paper we show that arrays of core-shell nanoparticles function as effective thin  absorbers of light. In contrast to known metamaterial absorbers, the introduced absorbers are formed by single planar arrays of spherical inclusions and enable full absorption of light incident on either or both sides of the array. We demonstrate possibilities for realizing different kinds of symmetric absorbers, including resonant, ultra-broadband, angularly selective, and all-angle absorbers. The physical principle behind these designs is explained considering balanced electric and magnetic responses of unit cells. Photovoltaic devices and thermal emitters are the two most important potential applications of the proposed designs.
\end{abstract}

\section{Introduction}
Recently, considerable research has been conducted on manipulating electromagnetic wave fronts using optically thin structures, so-called metasurfaces, which are two-dimensional arrays of sub-wavelength polarizable unit cells \cite{Huygens1,Huygens2,Transparency,Metamirrors}. Electromagnetic wave absorbers, realizing one of many possible functionalities of metasurfaces, have been also widely studied (see a review in \cite{Padilla}). 
A great deal of interest in creation of optically thin absorbers for infrared \cite{Ramakrishna} and visible \cite{Atwater,Yu} ranges has recently emerged in view of potential  applications in solar photovoltaic \cite{Polman} and thermo-photovoltaic \cite{TPV} systems. 
There have been many different designs proposed to realize such thin absorbers. 

However, most of the known realizations of optically thin absorbing layers require coherent illumination of \emph{both} sides of the layer. This can be achieved by positioning the absorbing layer close to a mirror reflector\cite{He,Fang,BackedCS1,BackedCS2,BackedCS3}, but  this apparently does not allow the harvesting of light energy from the space behind the mirror. Together with the reflector, such absorbers form asymmetric structures (with inevitable bianisotropic coupling \cite{absorption}), which can efficiently absorb electromagnetic waves hitting only one of their sides. Another drawback of reflector-backed absorbers is complete  blockage of  propagation of electromagnetic waves at all frequencies, outside of the absorption frequency band. Alternatively, coherent illumination of both sides of a lossy sheet can be realized using a system of beam splitters and mirrors, as in so-called \emph{coherent absorbers} \cite{coherent0,coherent1,coherent2}. This approach also does not allow full light harvesting in the general case when there are incoherent sources both in front and behind a thin absorbing sheet.  

In this paper we study a means to fully harvest energy of light illuminating either or both sides of a single-layer array of nanoparticles, independently of coherent or incoherent nature of light sources which can be arbitrarily located in space. To the best of our knowledge,  there have been only a few designs proposed to absorb electromagnetic waves illuminating a thin layer from one or both sides. Using a periodic racemic arrangements of right-handed and left-handed helices (so that the bianisotropy is compensated) one can create a symmetric absorbing layer \cite{Viicha}. However, this design is extremely challenging to realize in the optical range. Another symmetric absorber could be realized utilizing an array of dielectric-coated cylinders \cite{Costas_cylinders}. However, this design assumes perfectly conducting cylinders, it is polarization-sensitive and also difficult for practical realization. 

\section{Theory}
In our earlier work on generic thin absorbers \cite{absorption}, we theoretically explored all the possible ways to design symmetric and asymmetric absorbers made of periodic arrays of sub-wavelength polarizable unit cells. It was shown that in order to realize a symmetric absorber, balanced electric and magnetic responses must be present in each unit cell. When an incident wave impinges normally on an infinite array of electrically and magnetically polarizable uniaxial inclusions, electric and magnetic dipole moments are induced in the unit cells. The corresponding induced electric and magnetic surface currents create two secondary plane waves, in the forward and backward directions. The amplitudes of these waves can be written as \cite{absorption}
\e\label{eq:A} 
E_{\rm forw}=\frac{i\omega}{2S}\left(\eta_0 p +m\right),\quad E_{\rm
back}=\frac{i\omega}{2S}\left(\eta_0 p -m\right),\f
where $p$, $m$, $\eta_0$, $\omega$, and $S$ are the electric moment, magnetic moment, free-space wave impedance, angular frequency, and the unit cell area, respectively. To ensure that the incident wave is fully absorbed in the layer, the array should be tuned in such a way that the forward-scattered wave cancels the  incident wave $E_{\rm forw}=-E_{\rm inc}$ (to make the transmission zero) and the backward-scattered wave has zero amplitude $E_{\rm back}=0$ (working as a Huygens' layer \cite{absorption,Samofalov}). This requires the  induced electric and magnetic moments to be equal to
\e\label{eq:B}  
p=\widehat{\alpha}_{\rm ee}E_{\rm inc},\qquad m=\eta_0 p=\widehat{\alpha}_{\rm mm}H_{\rm inc}\f
in which the required effective electric polarizability $\widehat{\alpha}_{\rm ee}$ and the effective magnetic polarizability $\widehat{\alpha}_{\rm mm}$ of the unit cells read
\e \label{eq:BR1} 
\frac{\eta_0}{S}\widehat{\alpha}_{\rm ee}=\frac{1}{\eta_0S}\widehat{\alpha}_{\rm mm}=\frac{i}{\omega}.
\f
To design a metasurface it is more preferable to work with  the polarizabilities of individual unit cells in free space instead of the effective polarizabilities of the same cells arranged in a periodical lattice. For a uniaxial periodic metasurface, the individual polarizabilities $\alpha_{\rm ee}$ and $\alpha_{\rm mm}$ in terms of the  effective polarizabilities can be written as $ 1/\alpha_{\rm ee}=1/\widehat{\alpha}_{\rm ee}+\beta_{\rm e}$, $1/\alpha_{\rm mm}=1/\widehat{\alpha}_{\rm mm}+\beta_{\rm m}$,
where $\beta_{\rm e}$ and $\beta_{\rm m}=\beta_{\rm e}/\eta_0^2$ are the interaction constants between the unit cells in the array \cite{Teemu}. Considering the requirements for symmetric full absorption in Eq.~(3), we can conclude that to fully absorb normally incident plane waves in an optically thin layer, the individual electric and magnetic polarizabilities should be balanced,  satisfying the following condition: 
\e \label{eq:C}
\eta_0\alpha_{\rm ee}=\frac{1}{\eta_0}\alpha_{\rm mm}=\left(\frac{\omega}{iS}+\frac{\beta_{\rm e}}{\eta_0}\right)^{-1}.\f
Knowing the desired polarizabilities of the unit cells given by Eq.~(4), we can proceed with designs of appropriate topologies of absorbing metasurfaces.

\begin{figure*}[t]
\centering
\subfigure[]{\includegraphics[width=0.3\textwidth]{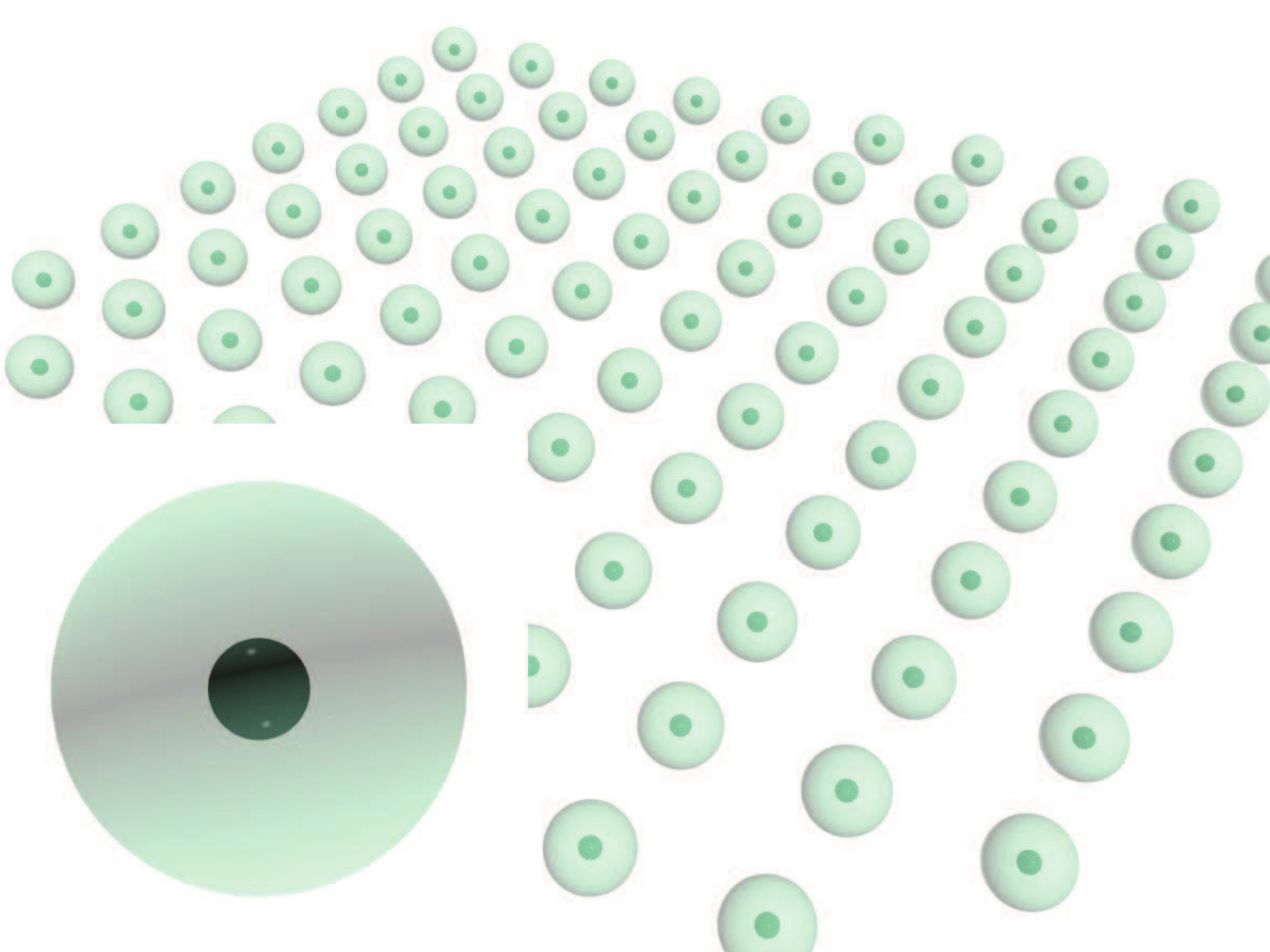} \label{N-Schem}}
\subfigure[]{\includegraphics[width=0.28\textwidth]{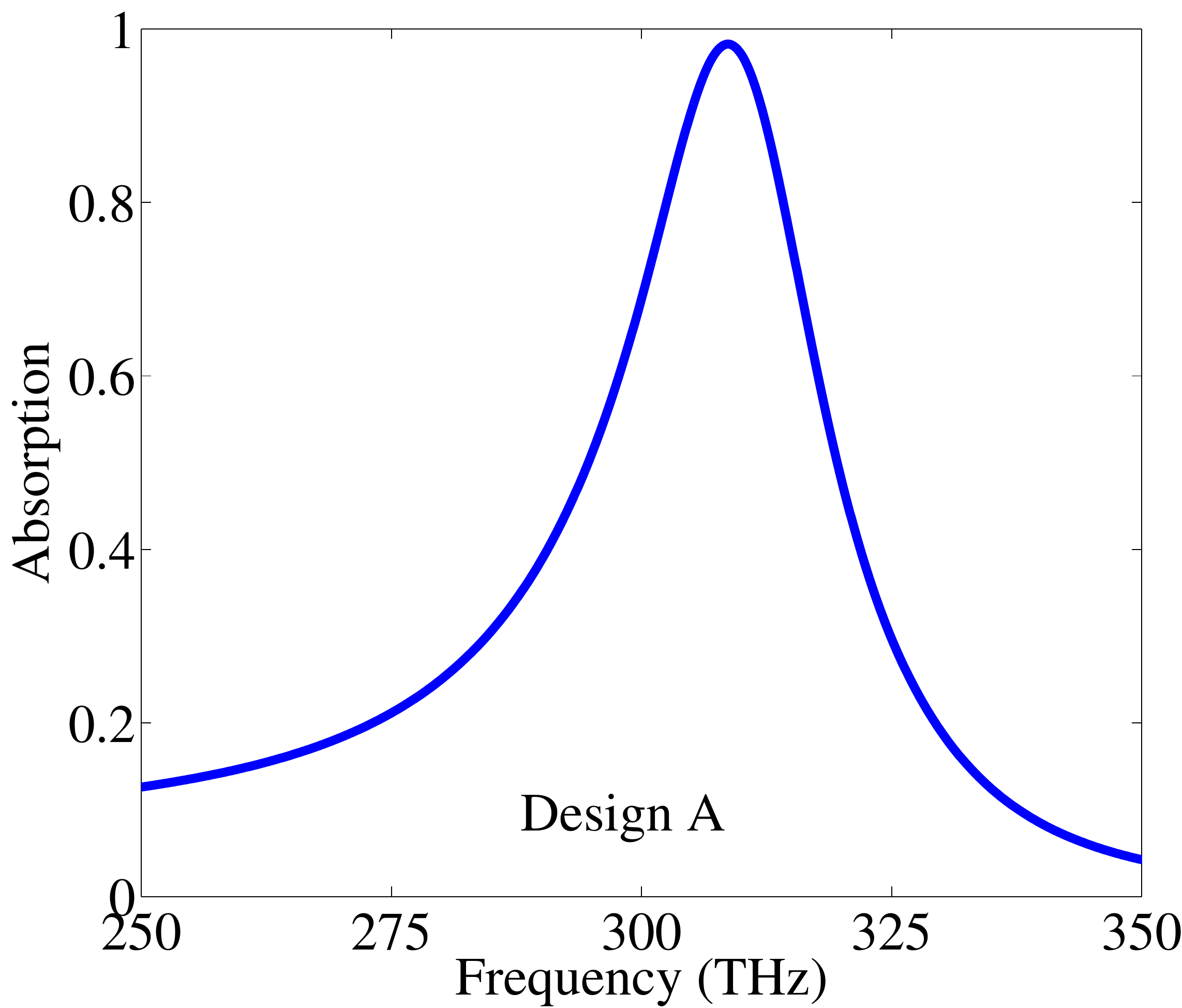} \label{N-A}}\\
\subfigure[]{\includegraphics[width=0.3\textwidth]{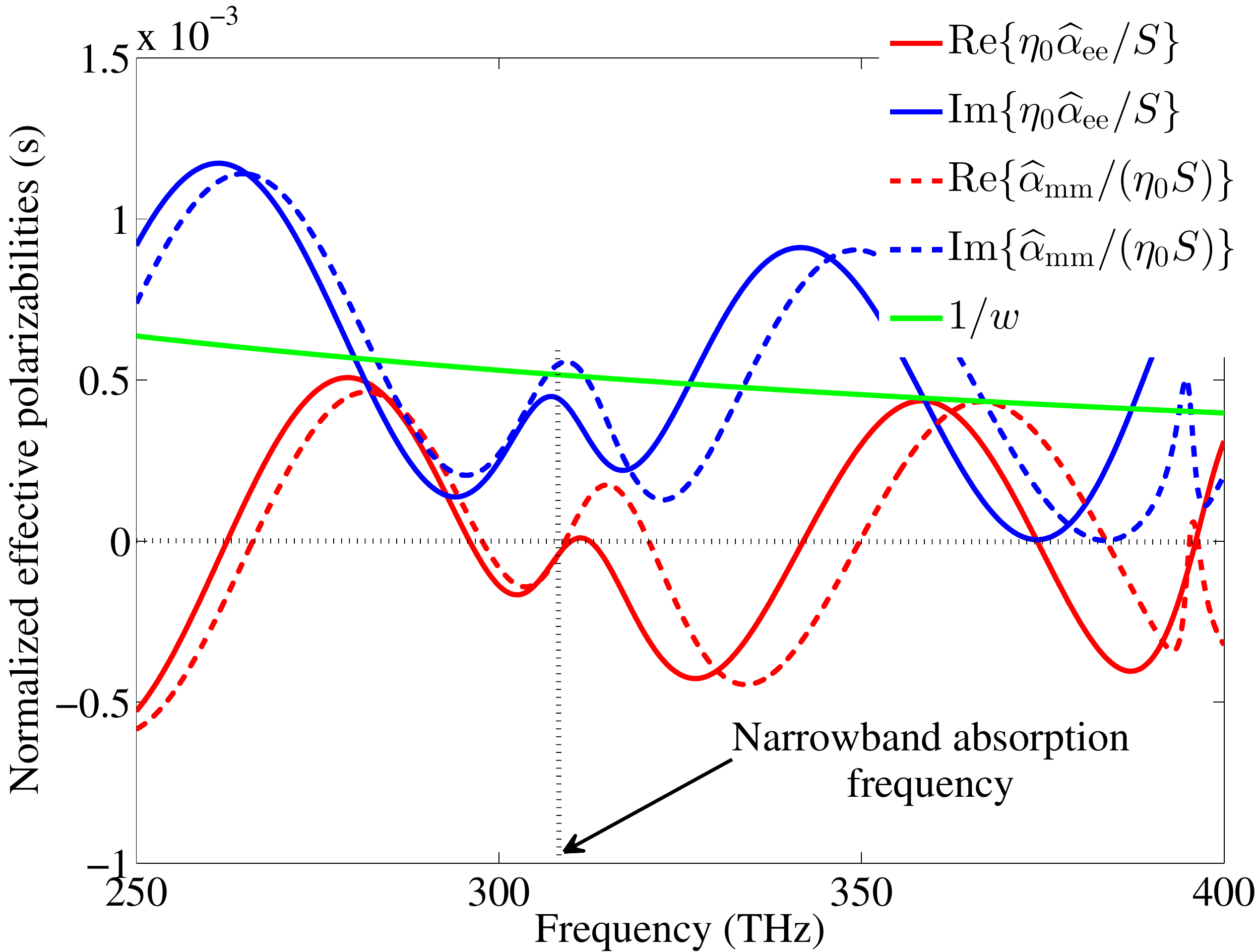} \label{EPolar-A}}
\subfigure[]{\includegraphics[width=0.18\textwidth]{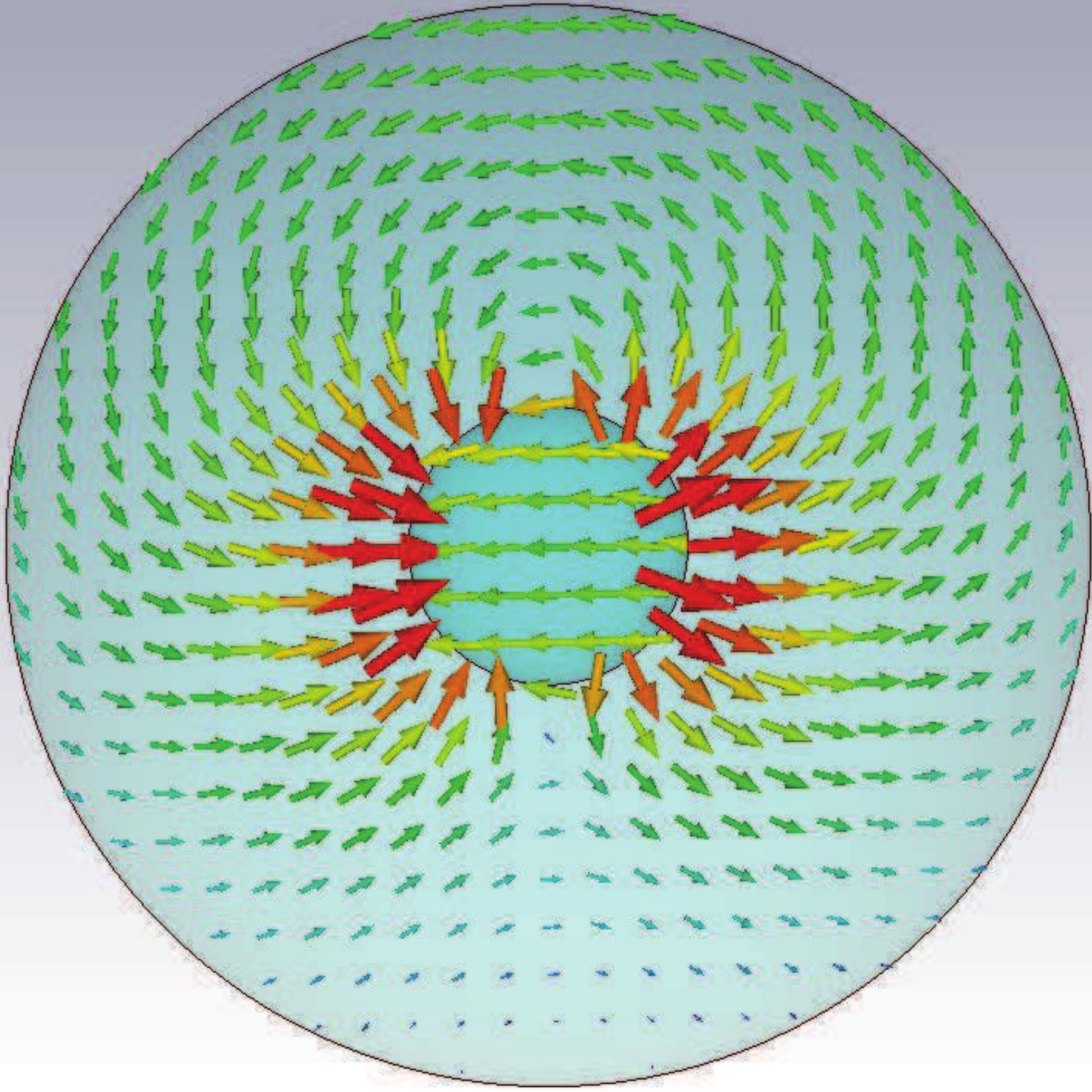} \label{N-D-E}}
\subfigure[]{\includegraphics[width=0.18\textwidth]{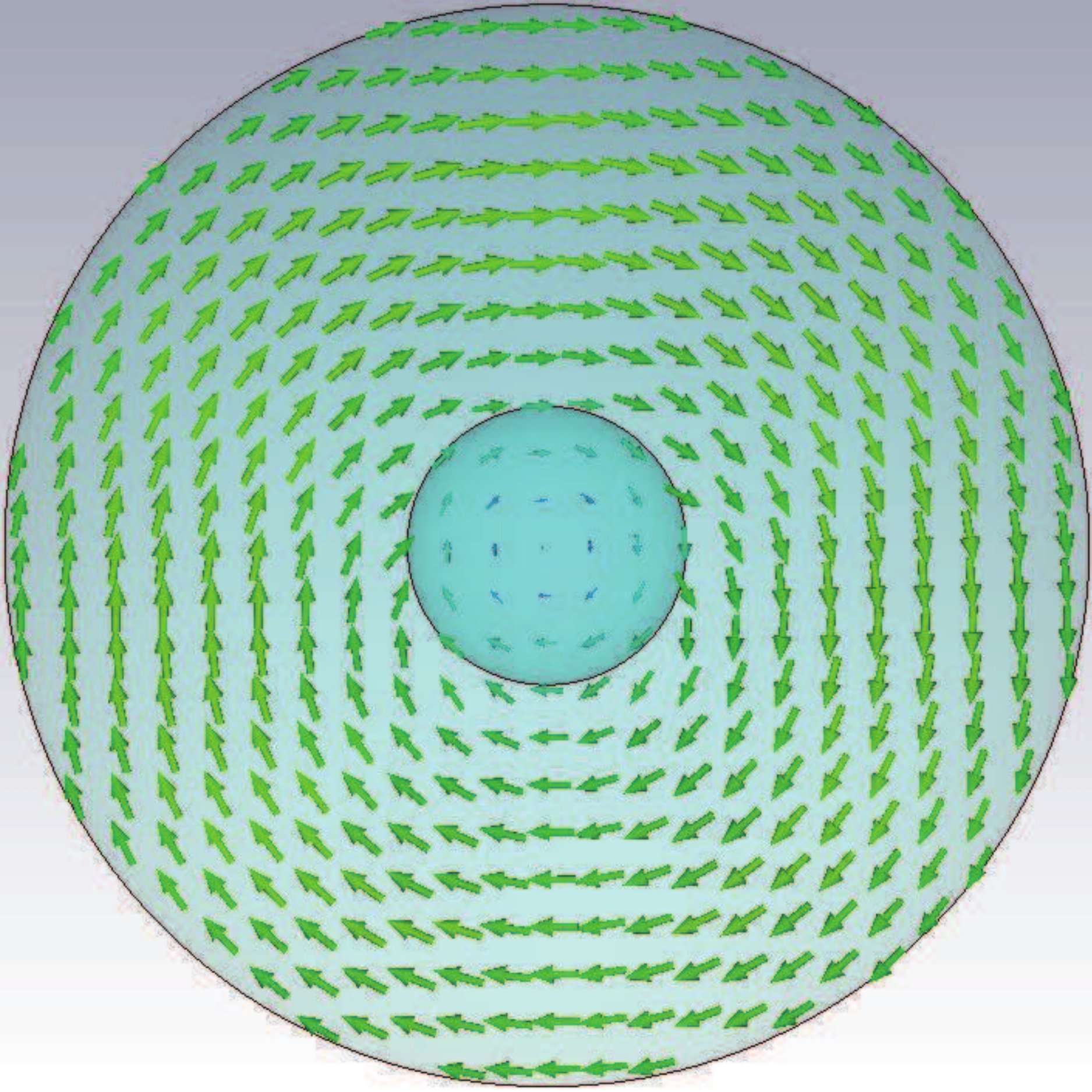} \label{N-D-M}}
\subfigure[]{\includegraphics[width=0.28\textwidth]{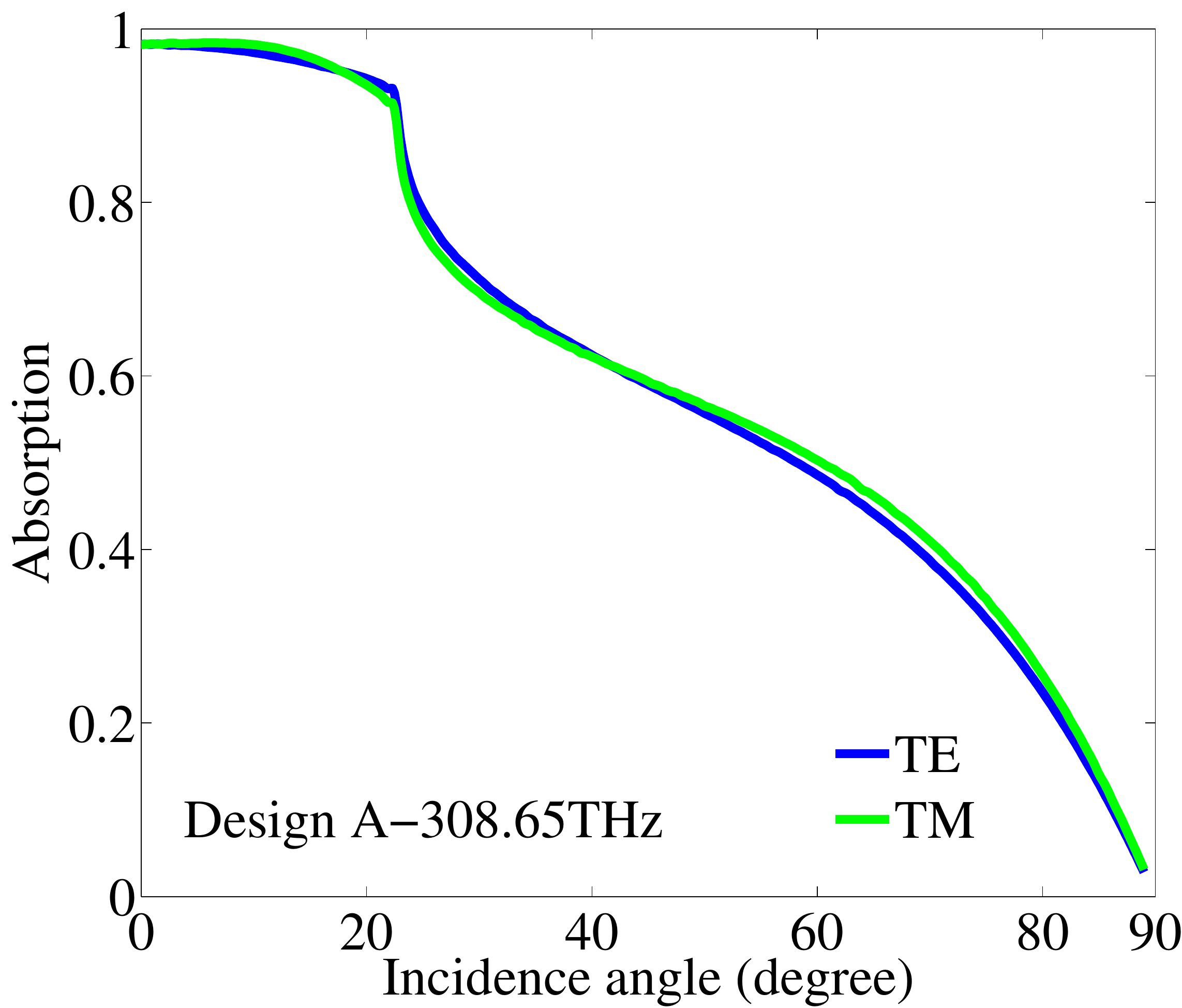} \label{N-AD}}
\caption{Design A: (a) Geometry of the proposed design. (b) Absorption as a function of the frequency. (c) Normalized effective electric and magnetic polarizabilities. (d) and (e) Electric current distributions with $\pi/2$ phase shift for 308.65 THz. (f) Absorption as a function of the incidence angle for 308.65 THz.}
\label{Design-A}
\end{figure*}

In this paper, we propose to utilize core-shell spherical nanoparticles, shown in Fig.~\ref{N-Schem}, as building blocks of symmetric light absorbers. This choice is motivated by the fact that similar particles were used previously to realize materials with negative refractive index \cite{miss,Morits}. In that application the electric and magnetic responses also need to be balanced, if it is desired to realize equal relative permittivity and permeability, but the absorption loss in the particles degrades the performance; in contrast, for our purpose the absorption loss becomes useful and we will show that a single array of properly tuned nanoparticles can fully absorb the incident light. Utilizing core-shell particles in light absorbers is not a new idea \cite{BackedCS1,BackedCS2,BackedCS3}. However, in all known designs an array of core-shell particles is located over a continuous metallic sheet. The presence of a metallic sheet in these structures is believed to be necessary for making transmission zero. Here we will show that there is no necessity for a metallic ground plane in absorber structures and a single array of core-shell particles can fully absorb light energy. This concept opens up a route to designing absorbers which can absorb light hitting either of their sides (symmetric structures lead to symmetric absorbers) while being transparent for incident waves out of the absorption frequency band.

For an optically small isotropic core-shell sphere, the electric and magnetic polarizabilities read 
\e\label{eq:D} 
\eta_0\alpha_{\rm ee}=\frac{6\pi  c^2}{i\omega^3}a_1,\quad\frac{1}{\eta_0}\alpha_{\rm mm}=\frac{6\pi  c^2}{i\omega^3}b_1,\f
where $a_1$ and $b_1$ are the Mie coefficients (e.g., \cite{Doyle}) and  $c$ is the speed of light.
Equating these polarizabilities to the required ones in Eq.~(4) we can design the aimed absorbing metasurface. In our design, we use silver as the core and amorphous n-doped silicon as the shell. To obtain reliable results, we use measured data for the material parameters of both silver \cite{Johnson} and amorphous n-doped silicon \cite{Ulrich} at optical frequencies. At the first  stage, we estimate the required dimensions of the particles and the array period using the analytical expressions in Eqs.~(4) and (5) (which assume the point-dipole approximation of particle interactions in the array) and then optimize them utilizing a commercial electromagnetic software \cite{CST}.  

\section{Results and Discussion}
{\bf Design A.} Let us start with designing a resonant, narrowband symmetric absorber.  In the structure  shown in Fig.~\ref{N-Schem}, the radius of the silver core, the thickness of the amorphous n-doped silicon shell, and the array period are $r_{\rm Ag}=44$~nm, $t_{\rm n-Si}=127$~nm, and $d=700$~nm, respectively. The simulated absorption spectrum for this design at  the normal incidence is plotted in Fig.~\ref{N-A}. It can be seen that more than $98\%$ of power absorption is achieved at 308.65 THz. As it was explained before, to fully absorb the incident wave, balanced electric and magnetic responses should be present in the layer [see Eq.~(3)]. Knowing the reflection and transmission coefficients from the layer one can simply calculate the effective polarizabilities of the layer \cite{absorption}. Figure \ref{EPolar-A} shows that for the absorption frequency the normalized effective electric and magnetic polarizabilities are balanced satisfying the required condition in Eq.~(3). Inspection of electric current distributions in a unit-cell particle can also give insight into the physics behind this absorption phenomenon. Two electric current distributions in the E-plane (the plane containing the electric field vector, the direction of the wave propagation, and the center of a unit cell), with a $\pi/2$ phase shift, are shown in Figs.~\ref{N-D-E} and \ref{N-D-M}. They clearly confirm the presence of both electric and magnetic polarizations. Here the magnetic moment is artificially created through  circulating electric polarization currents in the E-plane. Since the nanoparticles are not optically very small, deviating incidence angle from the normal, one can expect higher order modes start to appear. Figure~\ref{N-AD} shows the power absorbed at  308.65 THz as a function of the incidence angle for both TE and TM polarizations. It is noticed that, with increasing incidence angle, absorption graph shows strong dispersion which is due to higher-order modes. 

\begin{figure*}[htbp]
\centering
\subfigure[]{\includegraphics[width=0.3\textwidth]{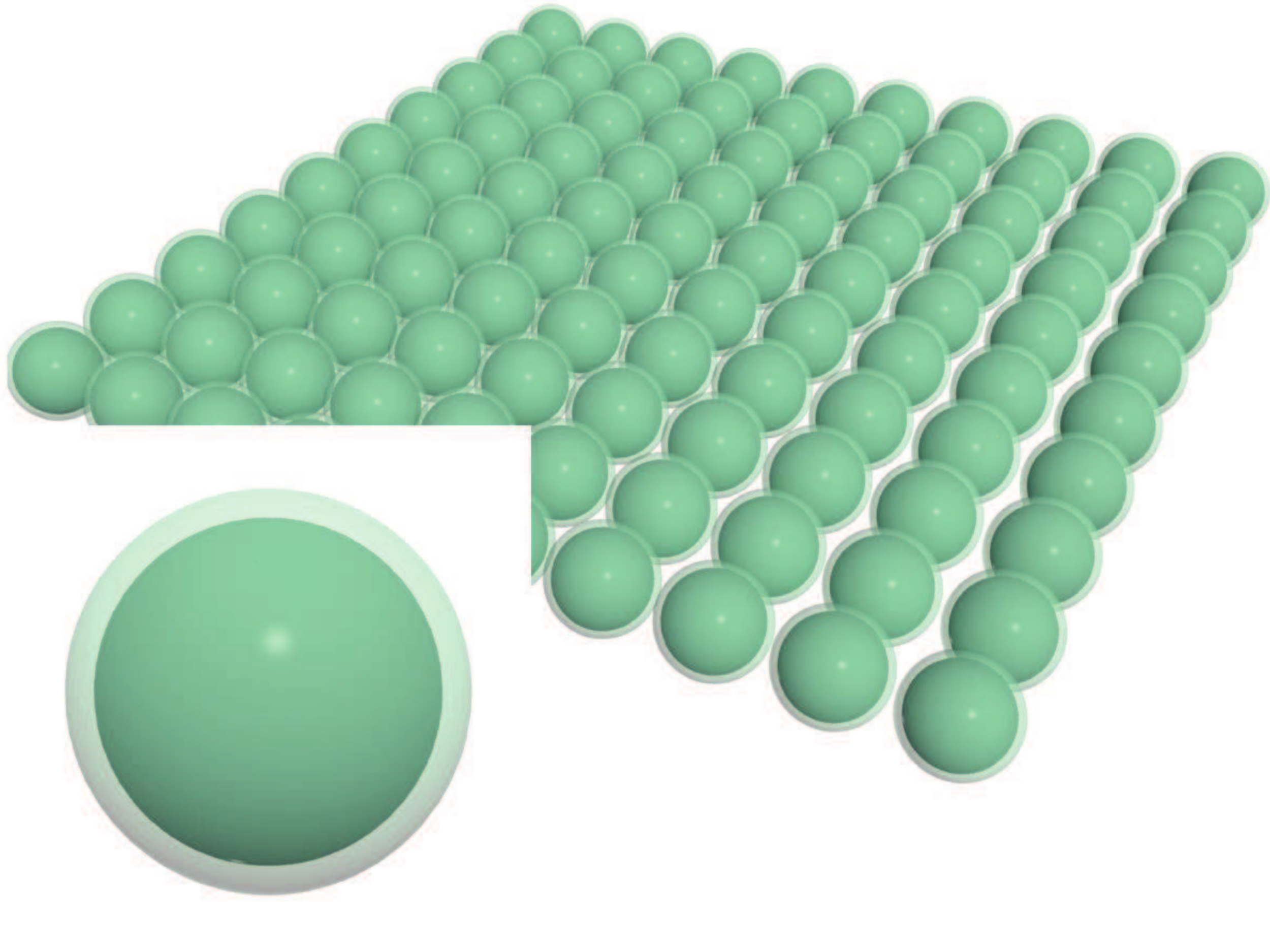} \label{B-Schem}}
\subfigure[]{\includegraphics[width=0.28\textwidth]{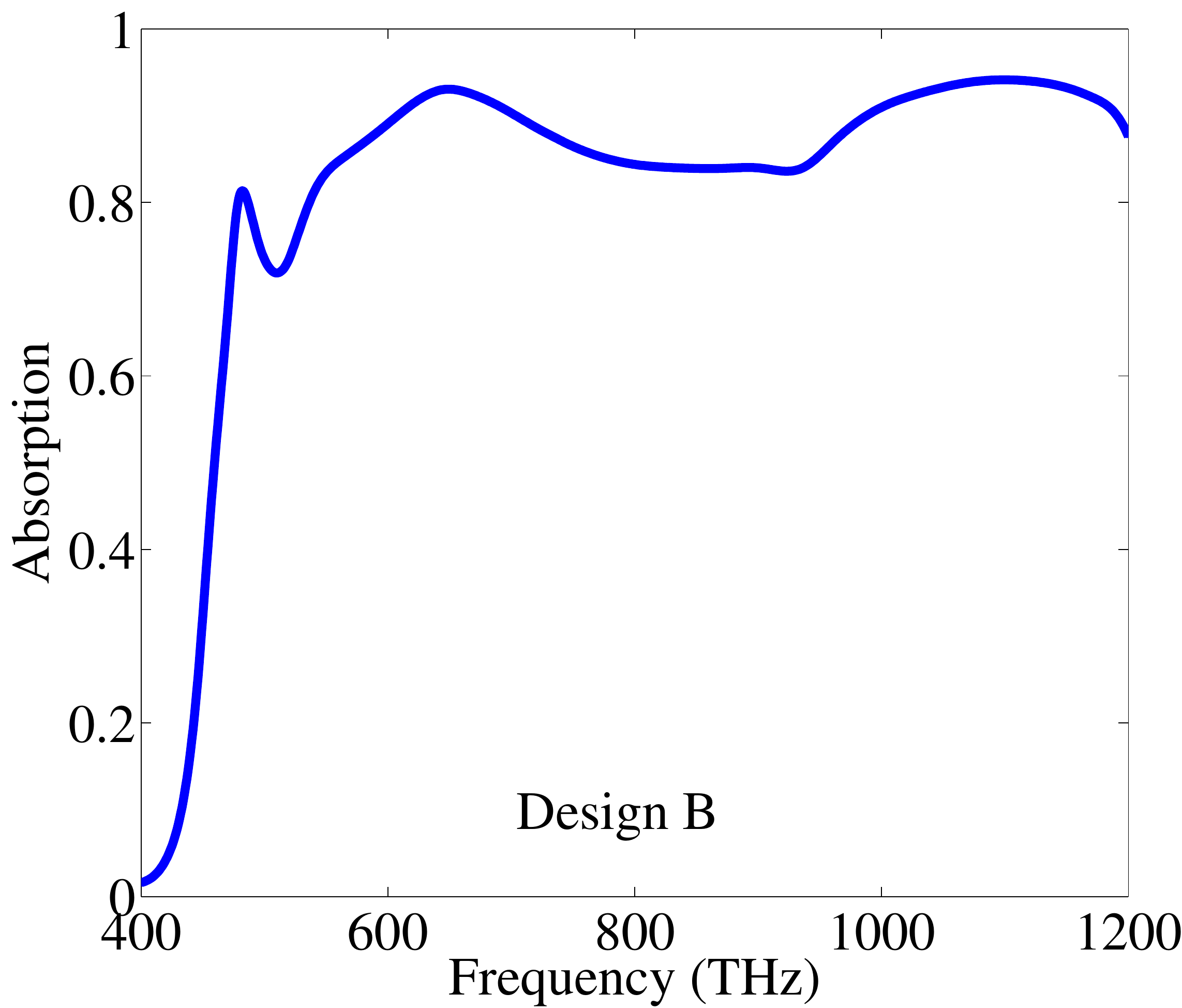} \label{B-A}}
\subfigure[]{\includegraphics[width=0.33\textwidth]{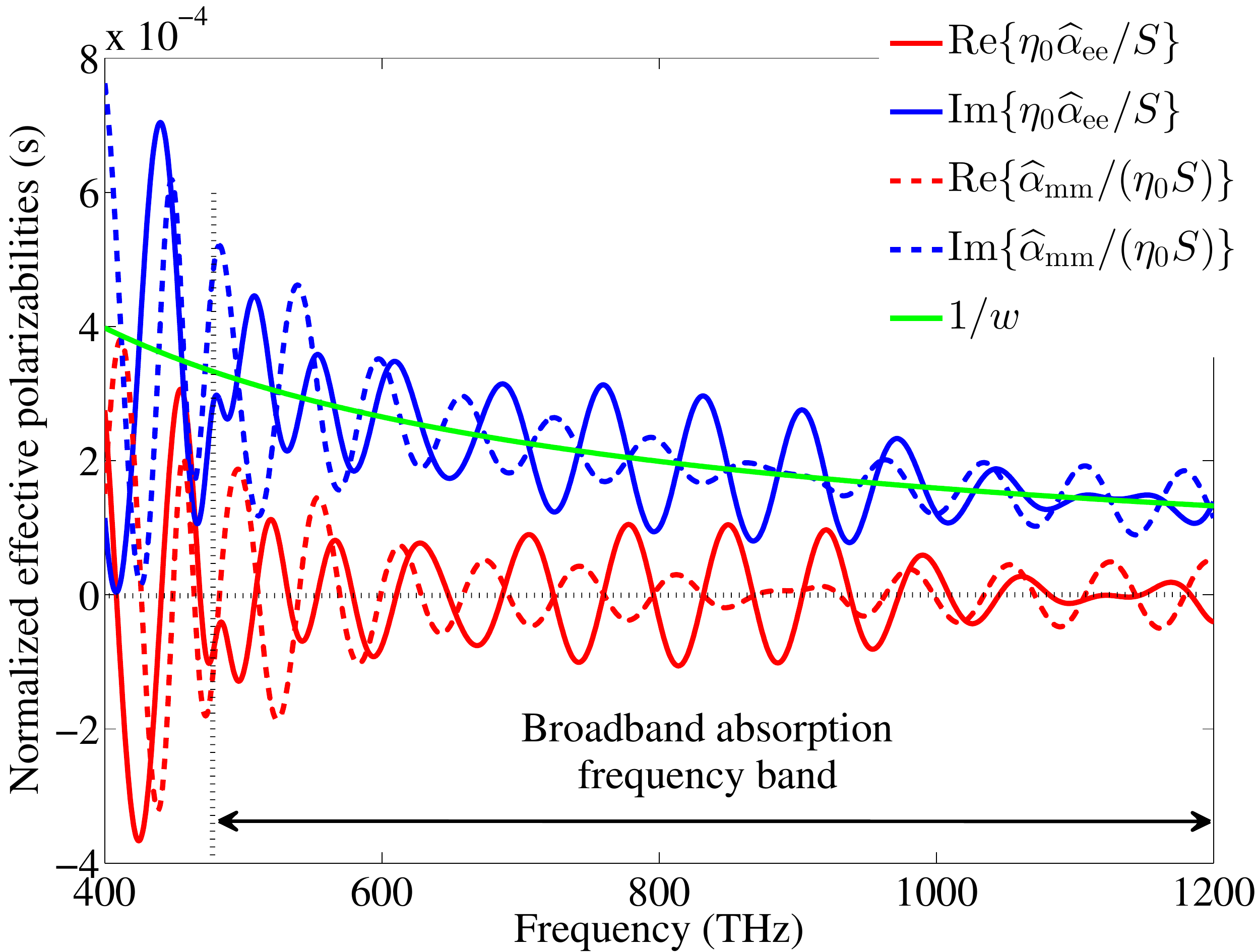} \label{EPolar-B}}
\subfigure[]{\includegraphics[width=0.18\textwidth]{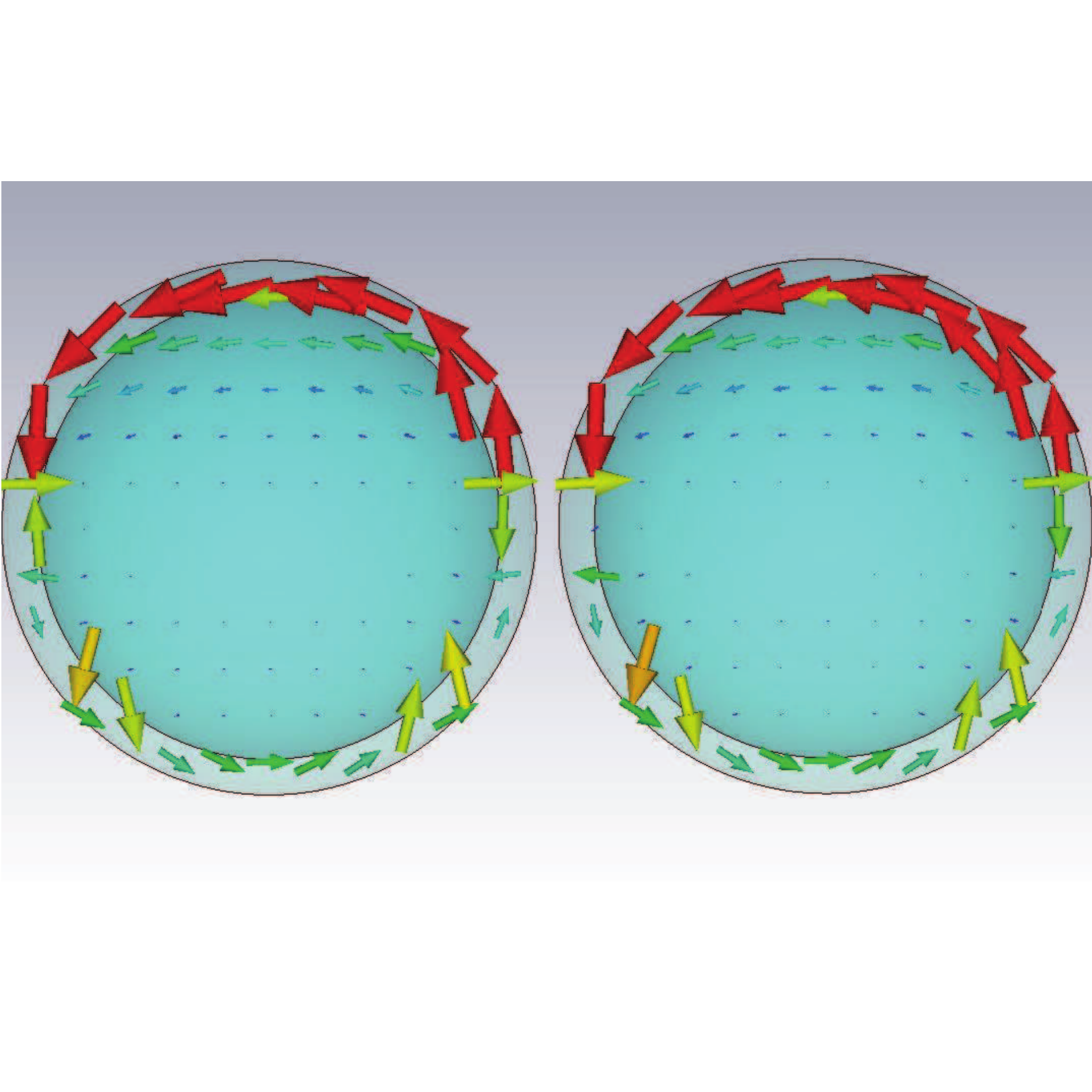} \label{B-D-E-600THz}}
\subfigure[]{\includegraphics[width=0.18\textwidth]{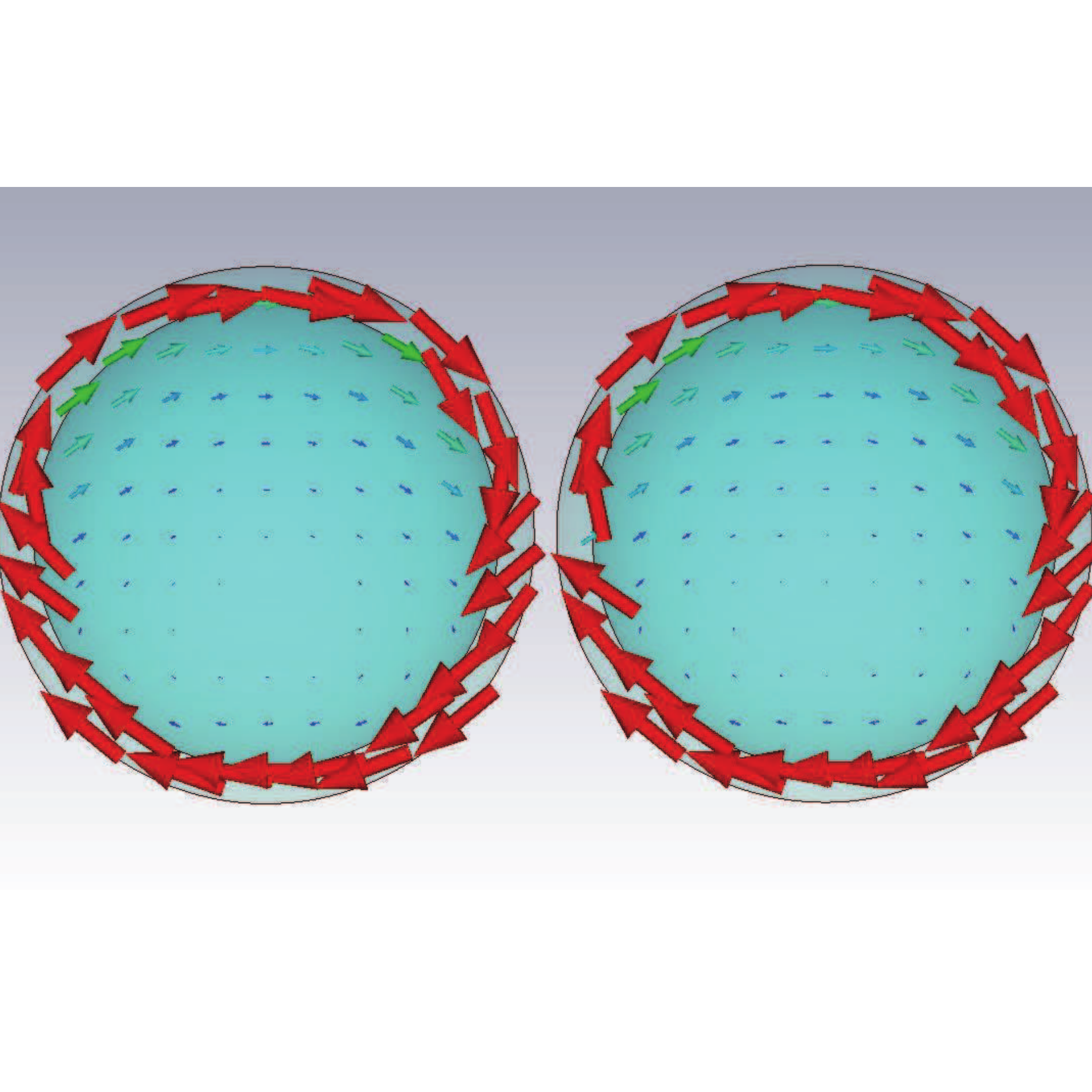} \label{B-D-M-600THz}}
\subfigure[]{\includegraphics[width=0.18\textwidth]{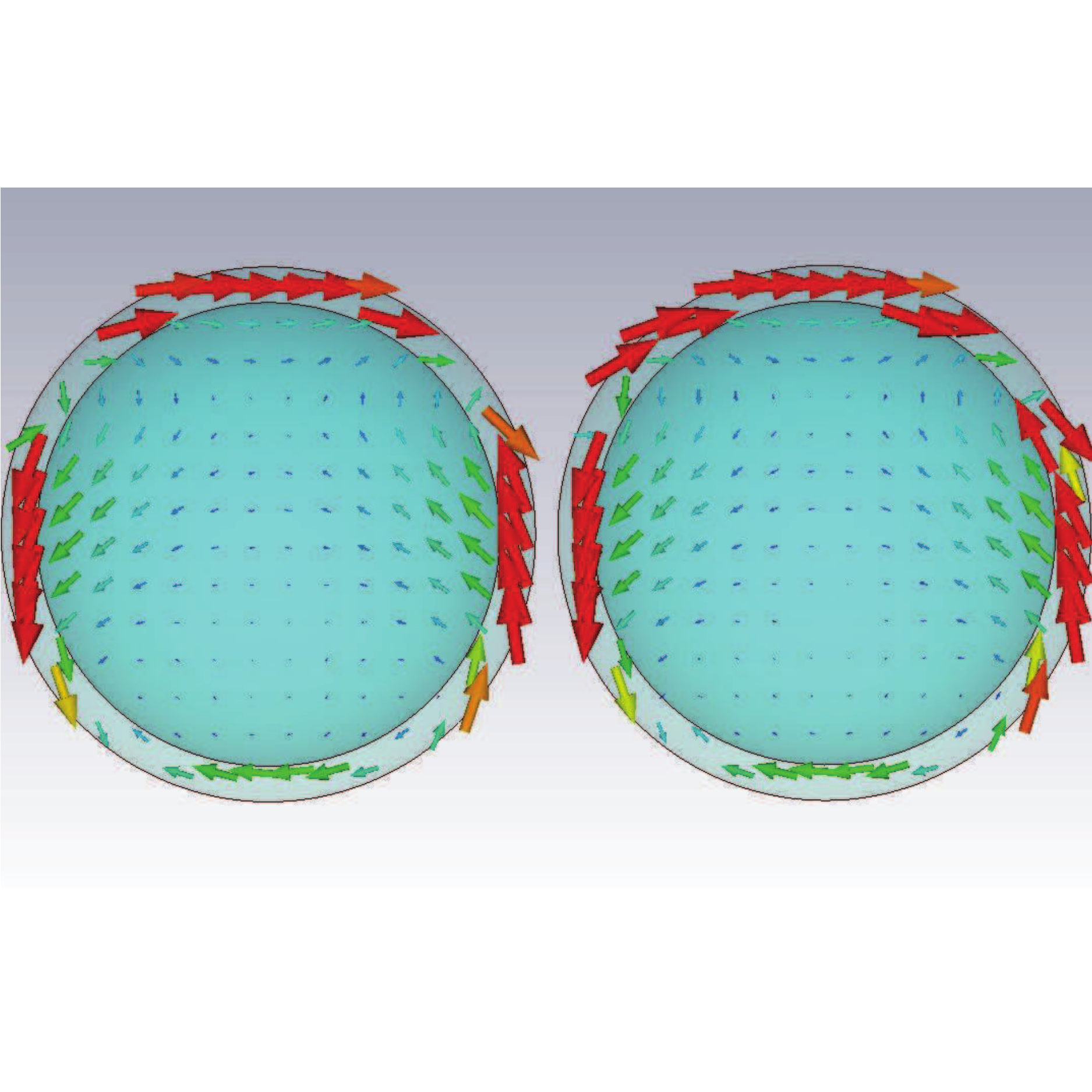} \label{B-D-E-1000THz}}
\subfigure[]{\includegraphics[width=0.18\textwidth]{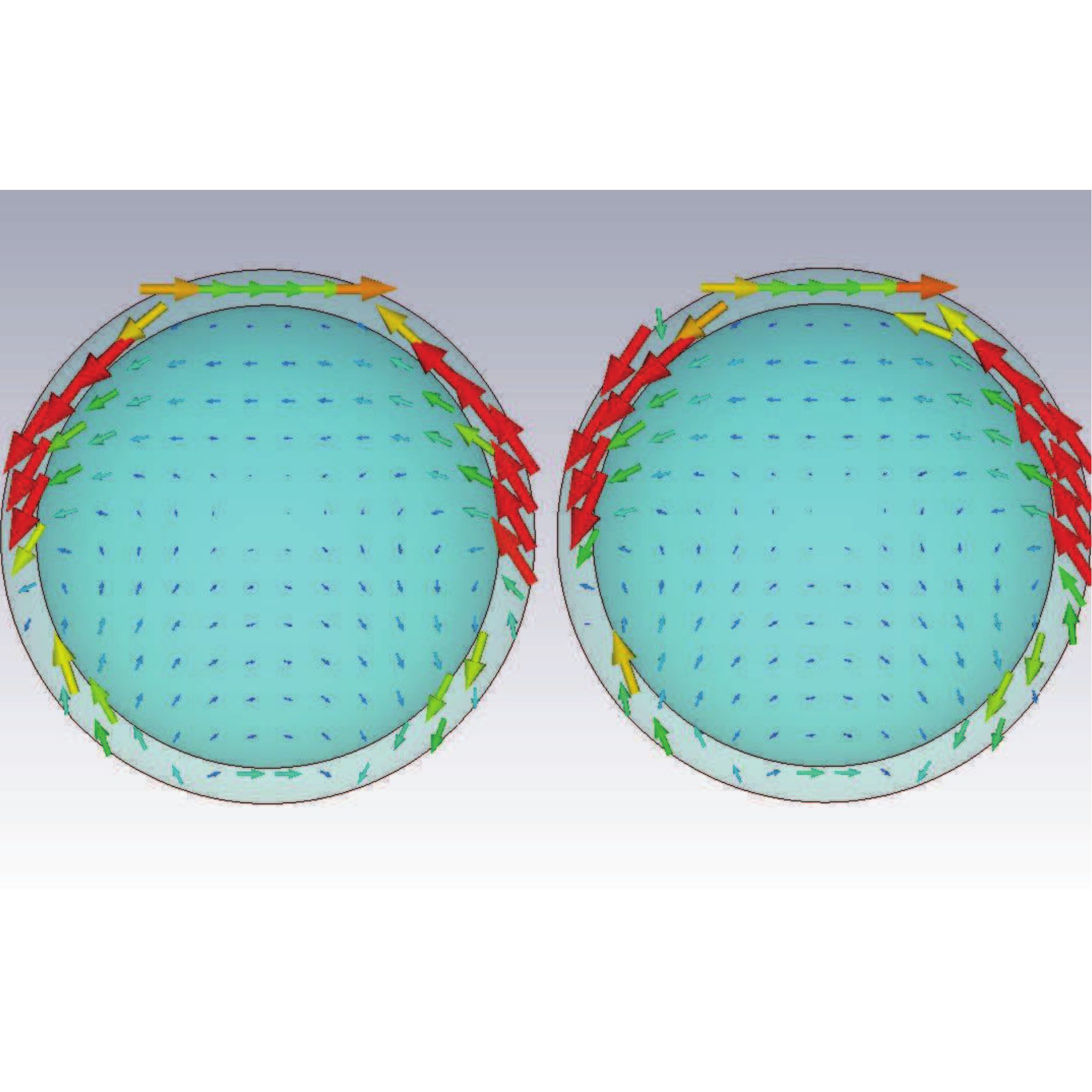} \label{B-D-M-1000THz}}\\
\subfigure[]{\includegraphics[width=0.23\textwidth]{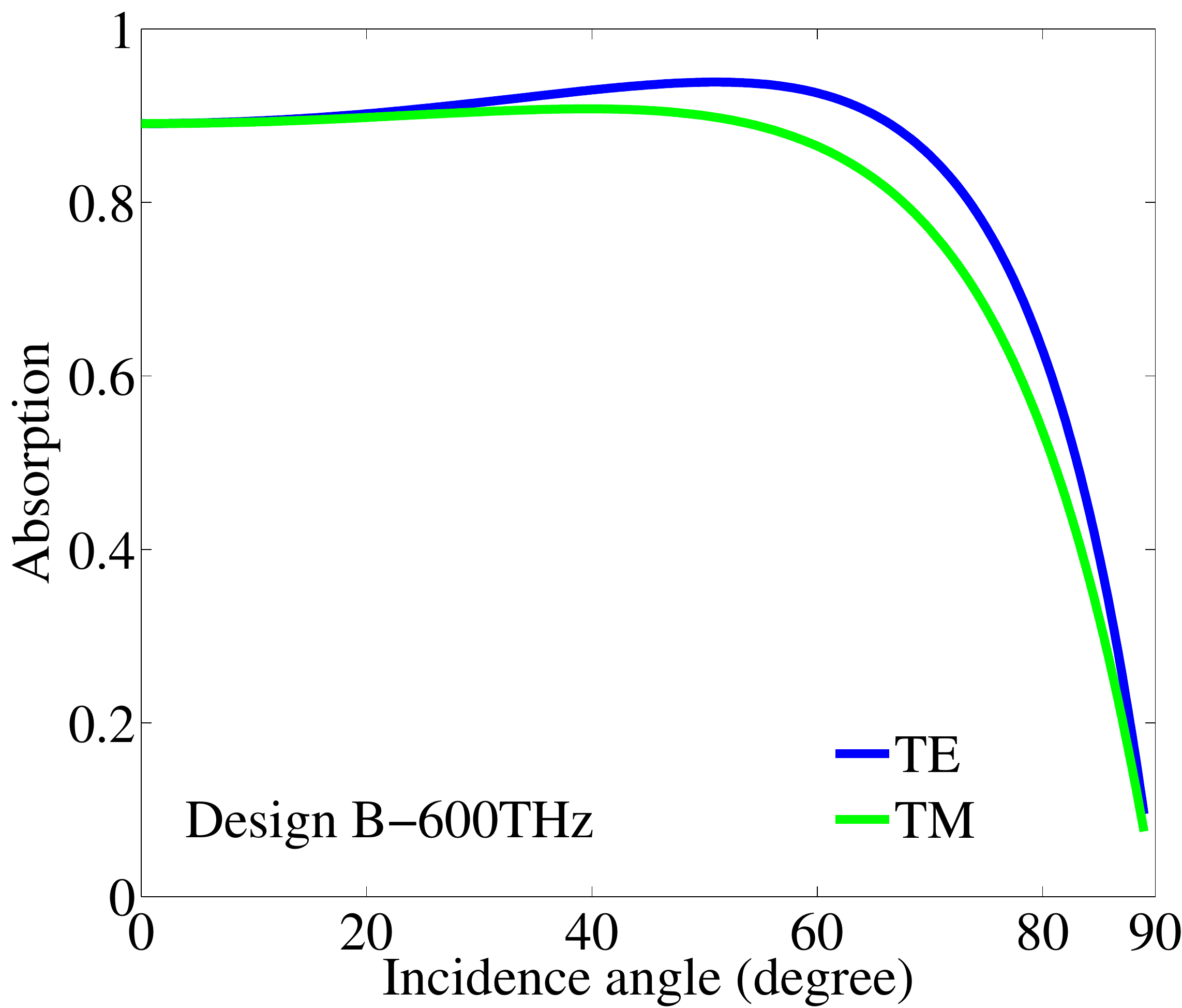} \label{B-AD-600THz}}
\subfigure[]{\includegraphics[width=0.23\textwidth]{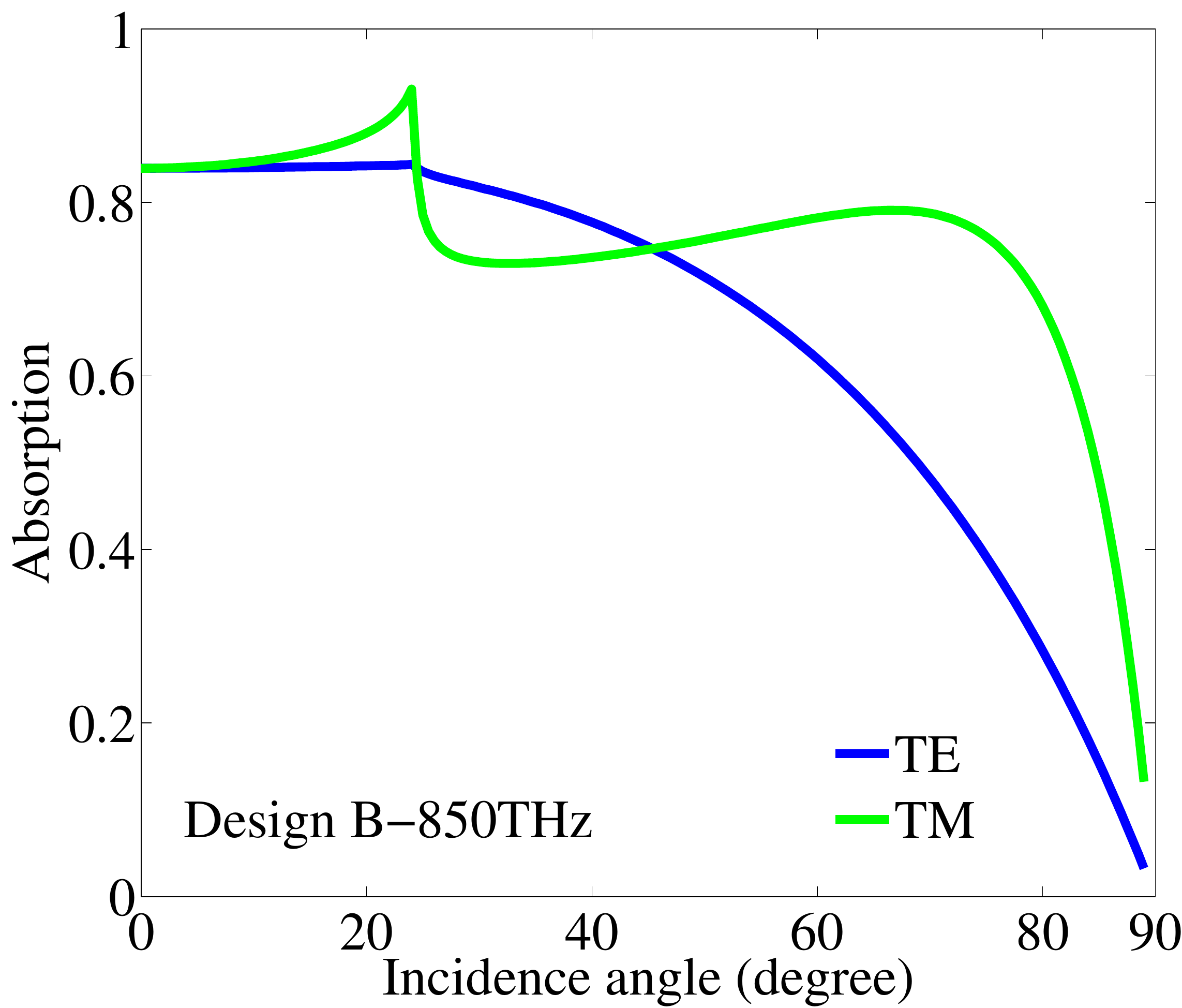} \label{B-AD-850THz}}
\subfigure[]{\includegraphics[width=0.23\textwidth]{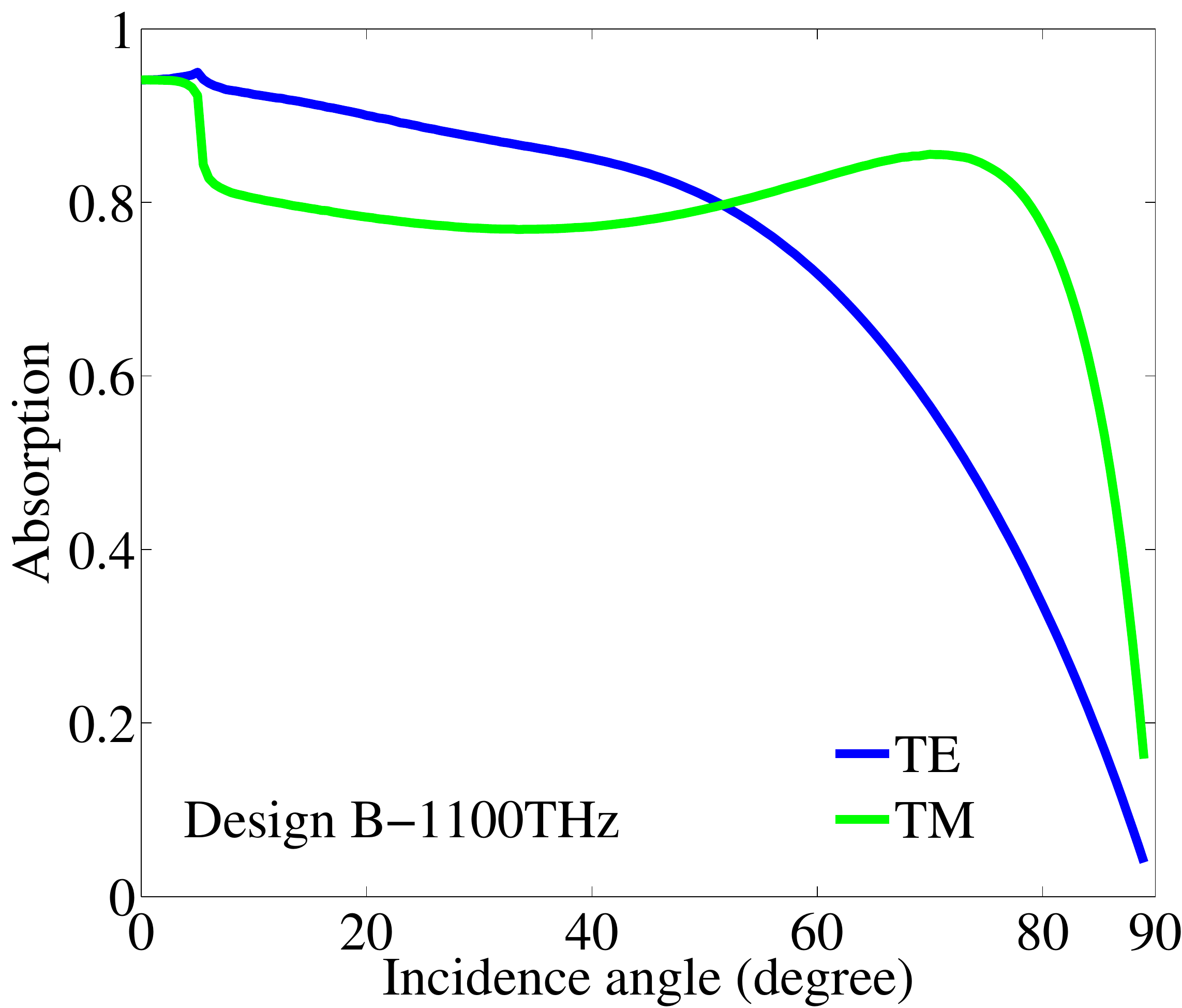} \label{B-AD-1100THz}}\\
\subfigure[]{\includegraphics[width=0.27\textwidth]{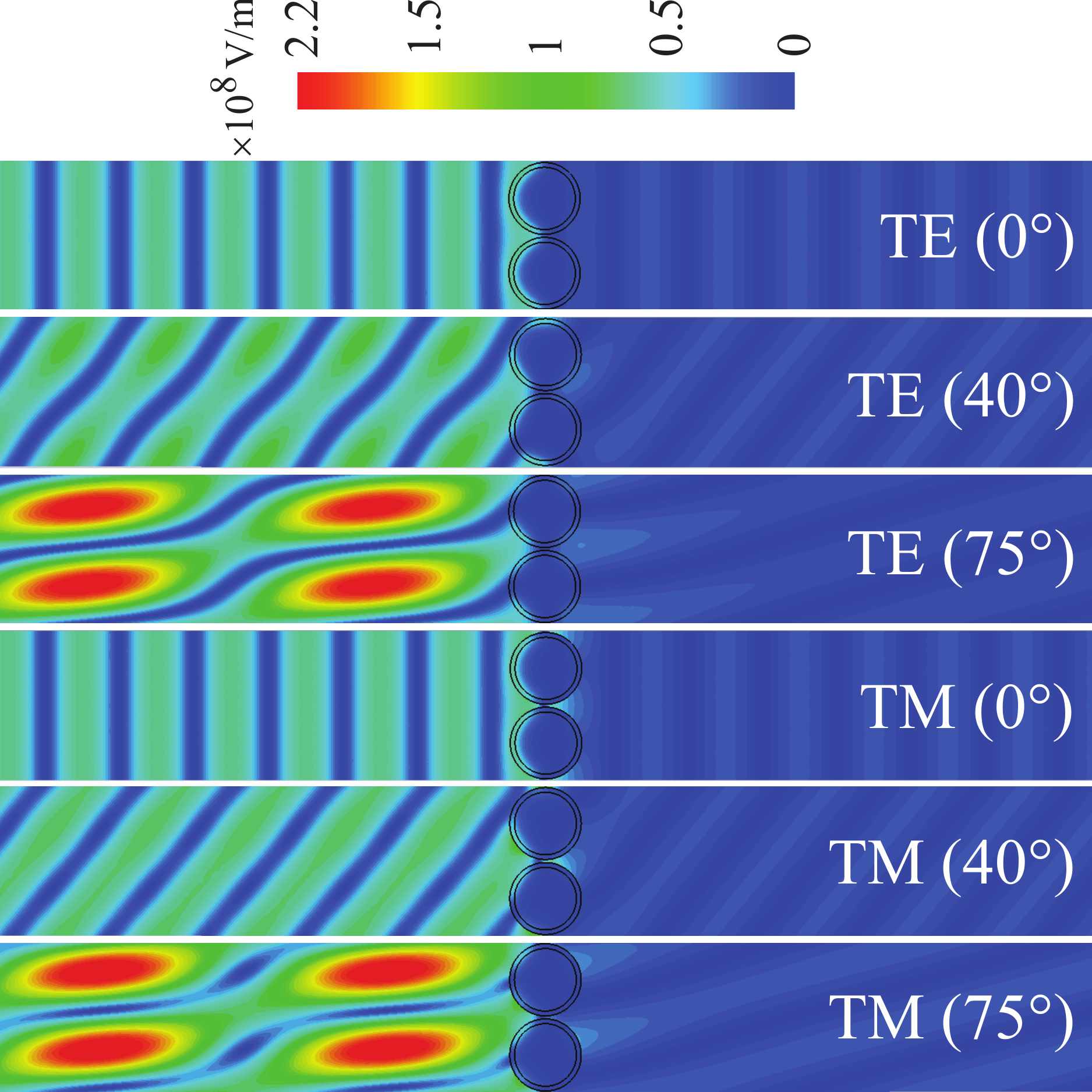} \label{B-W-600}}
\subfigure[]{\includegraphics[width=0.27\textwidth]{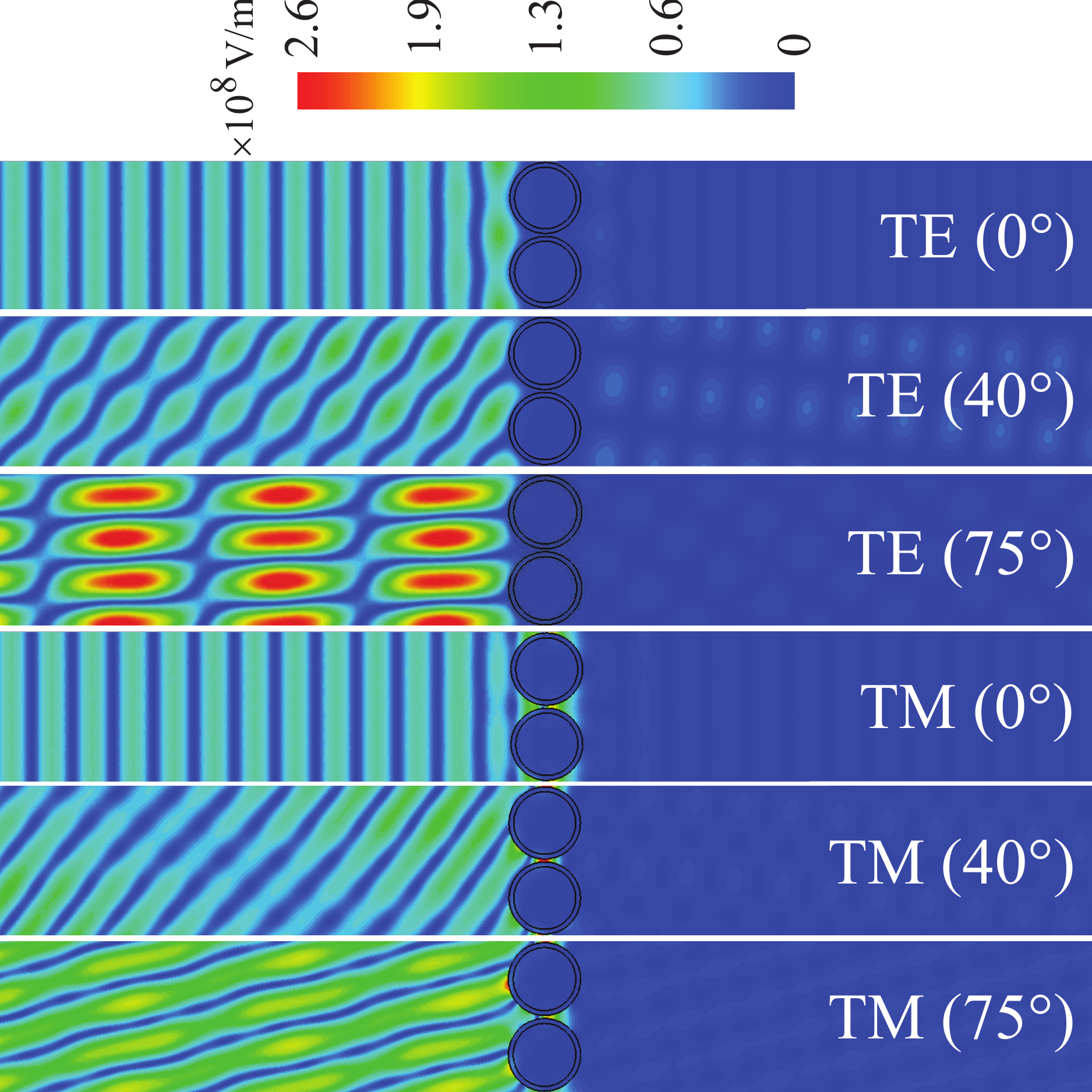} \label{B-W-1100}}
\caption{Design B: (a) Geometry of the proposed design. (b) Absorption as a function of the frequency. (c) Normalized effective electric and magnetic polarizabilities. (d) and (e) Electric current distributions with $\pi/2$ phase shift for 600 THz. (f) and (g) Electric current distributions with $\pi/2$ phase shift for 1000 THz. Absorption as a function of the incidence angle for (h) 600 THz, (i) 850 THz, and (j) 1100 THz. Electric field distributions (time-harmonic regime, field strength at a fixed moment of time is plotted) for TE (in the E-plane) and TM (in the H-plane) for (k) 600 THz and (l) 1100 THz.}
\label{Design-B}
\end{figure*}

{\bf Design B.} In the previous case we designed a narrowband resonant absorber. However, in applications such as energy harvesting it is of importance to design absorbers operating over a wide frequency range. One possibility to broaden the absorption bandwidth is to use an array made of particles with slightly different resonance frequencies \cite{Jokerst,Mayer}. This approach looks promising, but the operational spectrum can not be widened considerably because of the limited number of available different inclusions. 

Another method to design a broadband absorber is based on using a uniform array with unit cells made by stacking a number of resonators with different resonance frequencies \cite{He,Fang}. However, the design is challenging for manufacturing and in most of the cases the structure is optically thick. Here we propose an alternative, more general and yet simpler technique to design a symmetric broadband light absorber with significantly high angular stability. For an absorber layer made of a uniform array of particles, operating close to their resonance frequencies (e.g., the previous design), broadband absorption is not easy to achieve. However, if the layer is made of particles which are working far from their resonance frequency, it is in fact possible  to achieve high (although not perfect) absorption in a broad frequency range. To this end, we use  core-shell nanoparticles similar to the previous design but with $r_{\rm Ag}=104$ nm, $t_{\rm n-Si}=16$ nm, and $d=250$ nm (see Fig.~\ref{B-Schem}). The absorption graph for the layer made of such nanoparticles at the normal incidence, shown in Fig.~\ref{B-A}, confirms the ultra-wideband behavior for this design, i.e., more than 80\% of power absorption for the frequency range 540--1200 THz. The normalized effective electric and magnetic polarizabilities in Fig.~\ref{EPolar-B} shows that for the absorption region the real parts of the polarizabilities are close to zero while the imaginary parts are approaching the aimed value given by Eq.~(3). The electric current distributions, as a proof for the presence of magnetic response and its balance with electric one, are shown in Figs.~\ref{B-D-E-600THz} and \ref{B-D-M-600THz} for 600 THz and in Figs.~\ref{B-D-E-1000THz} and \ref{B-D-M-1000THz} for 1000 THz in a similar fashion to the previous case. An important point which can be seen from these distributions is that at higher frequencies the magnetic moment currents are mainly confined in between the adjacent nanoparticles instead of being distributed inside the inclusions. This promises a good angular stability for the structure. The results presented in Figs.~\ref{B-AD-600THz} (for 600 THz), \ref{B-AD-850THz} (for 850 THz), and \ref{B-AD-1100THz} (for 1100 THz) clearly indicate that the design proposed here exhibits a very good angular stability for different operational frequencies inside the broad absorption spectrum. Electric field distribution for TE and TM modes at different incidence angles for illuminations at 600 THz and 1100 THz are shown in Figs.~\ref{B-W-600} and \ref{B-W-1100}.

\begin{figure}[htbp]
\centering
\subfigure[]{\includegraphics[width=0.3\textwidth]{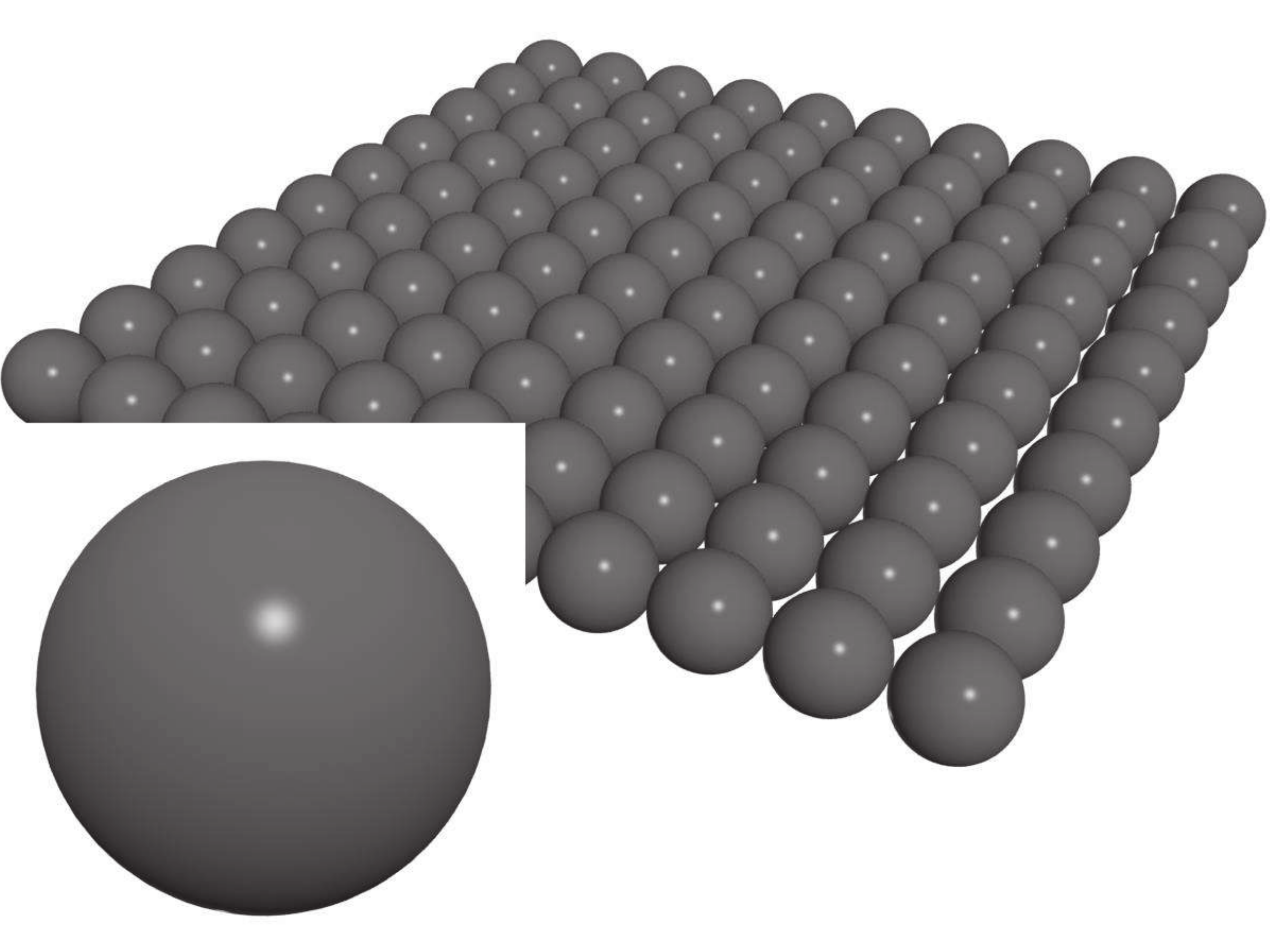} \label{AOS-Schem}}
\subfigure[]{\includegraphics[width=0.28\textwidth]{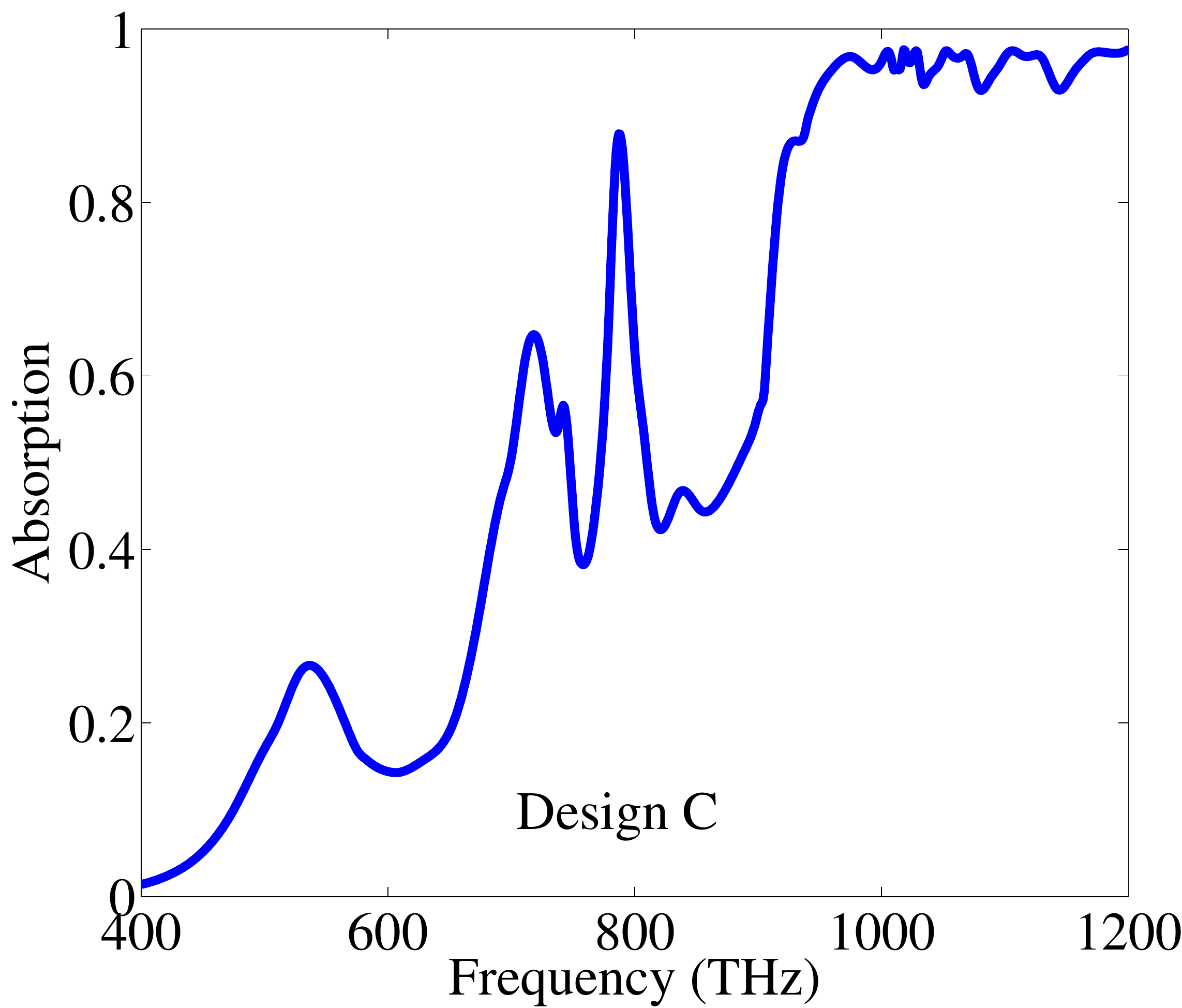} \label{AOS-A}}
\subfigure[]{\includegraphics[width=0.33\textwidth]{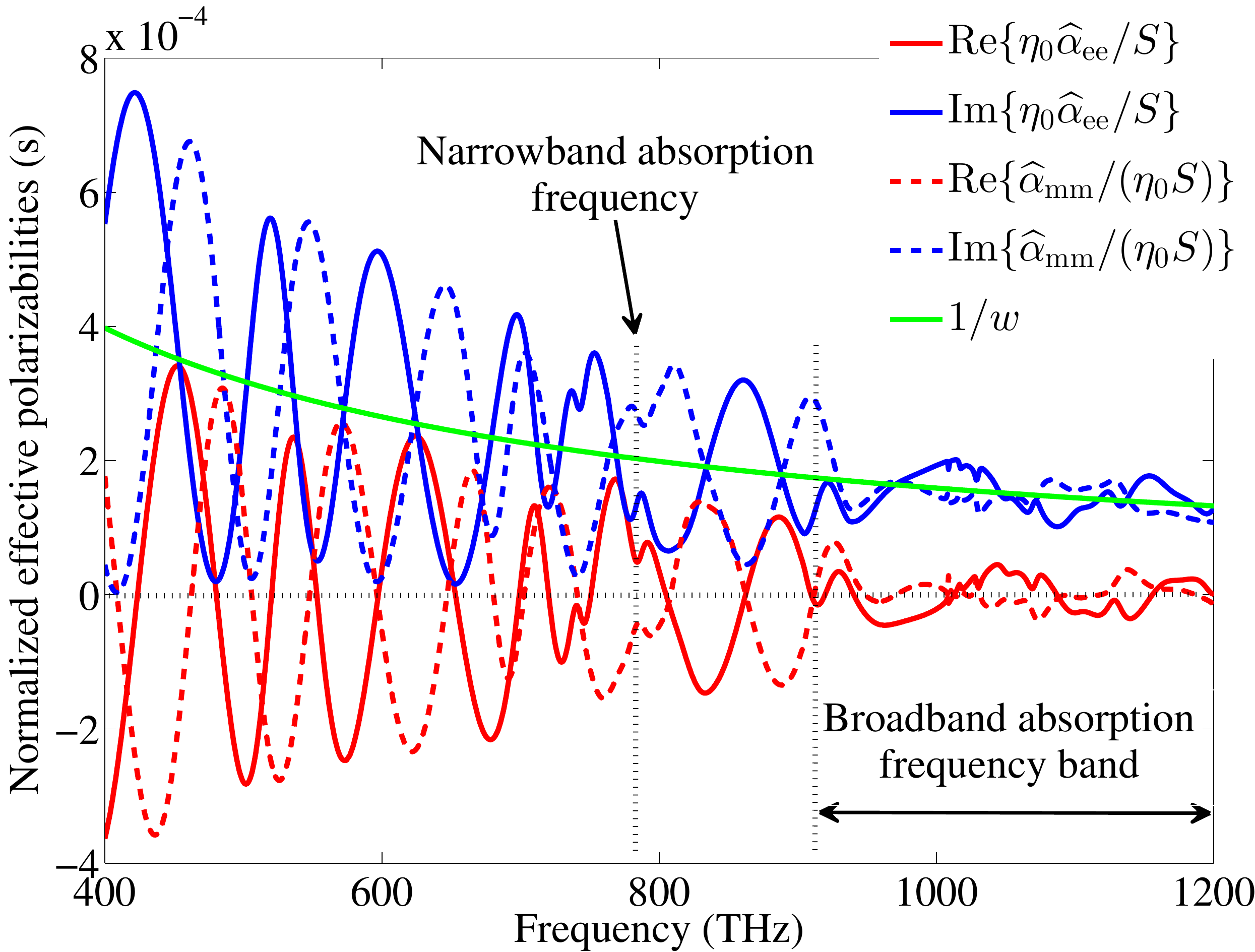} \label{EPolar-C}}\\
\subfigure[]{\includegraphics[width=0.18\textwidth]{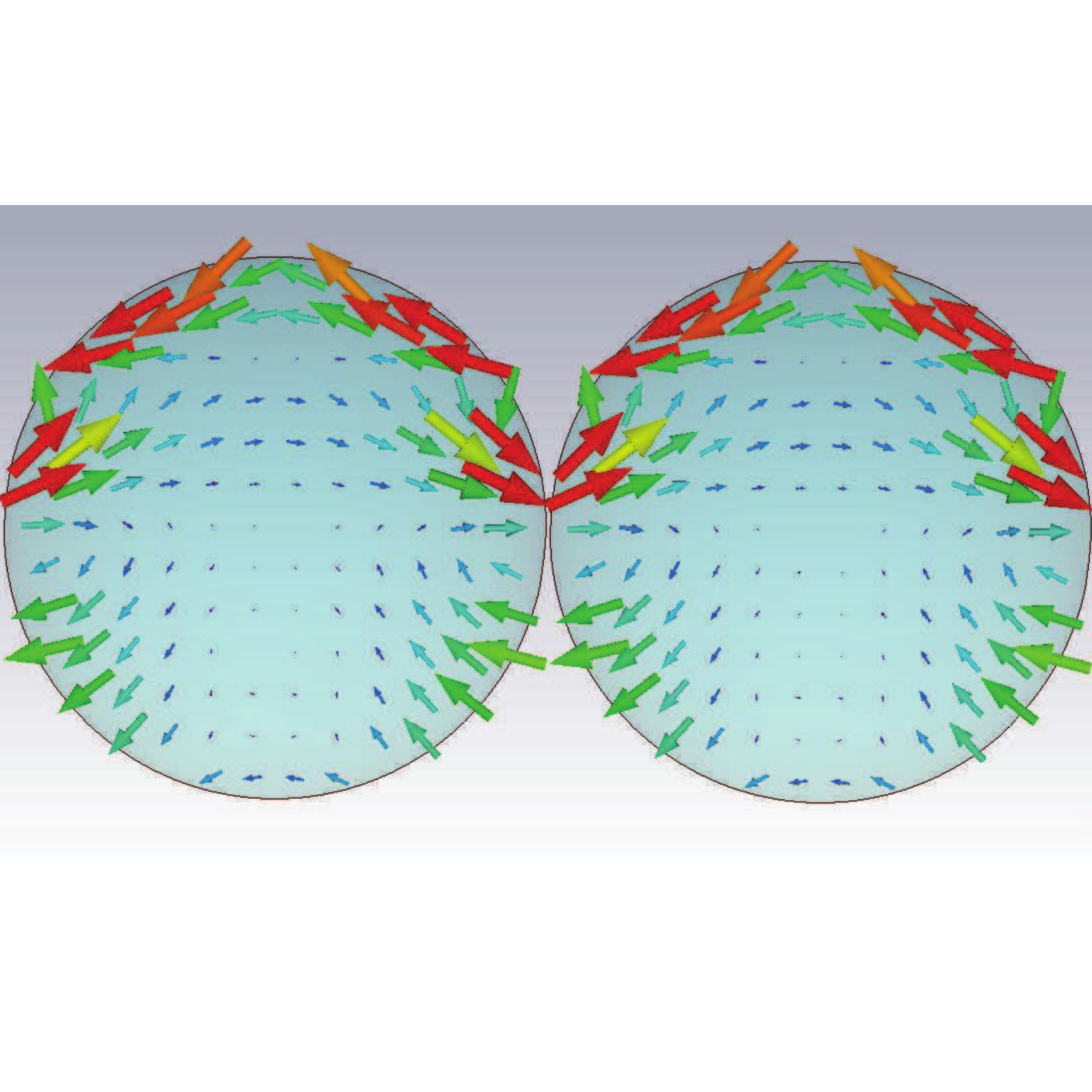} \label{AOS-D-E-1000THz}}
\subfigure[]{\includegraphics[width=0.18\textwidth]{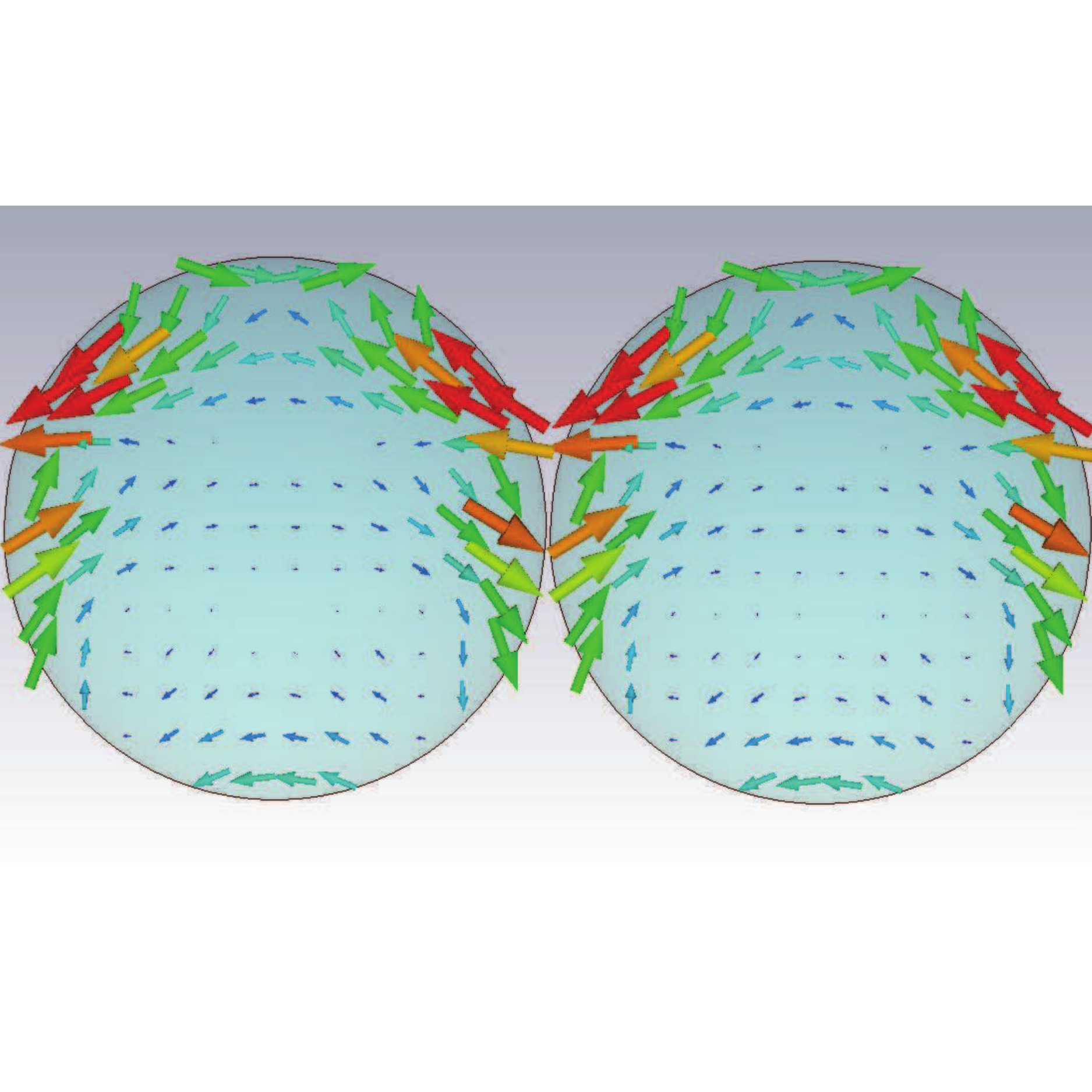} \label{AOS-D-M-1000THz}}\\
\subfigure[]{\includegraphics[width=0.23\textwidth]{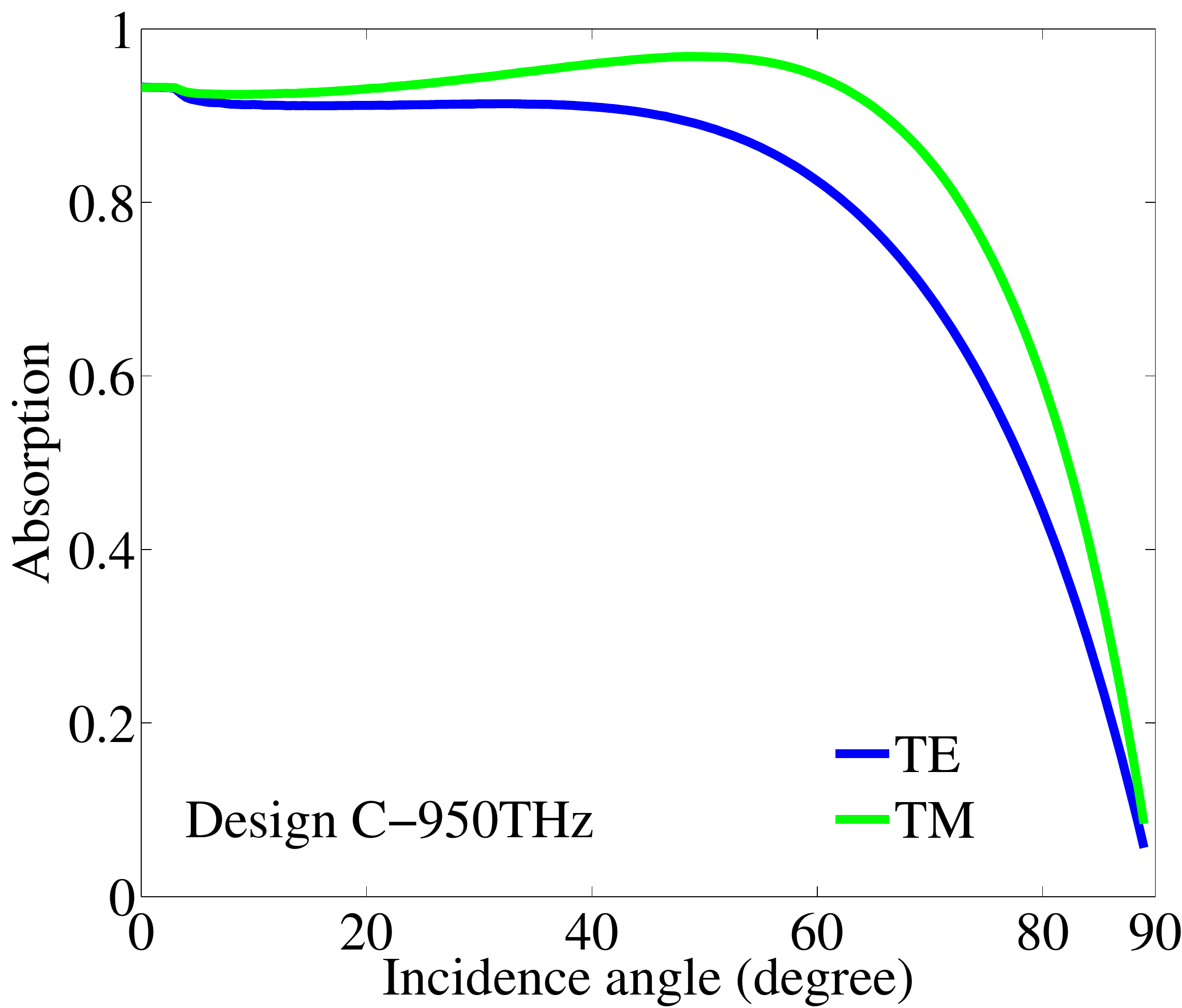} \label{AOS-AD-950THz}}
\subfigure[]{\includegraphics[width=0.23\textwidth]{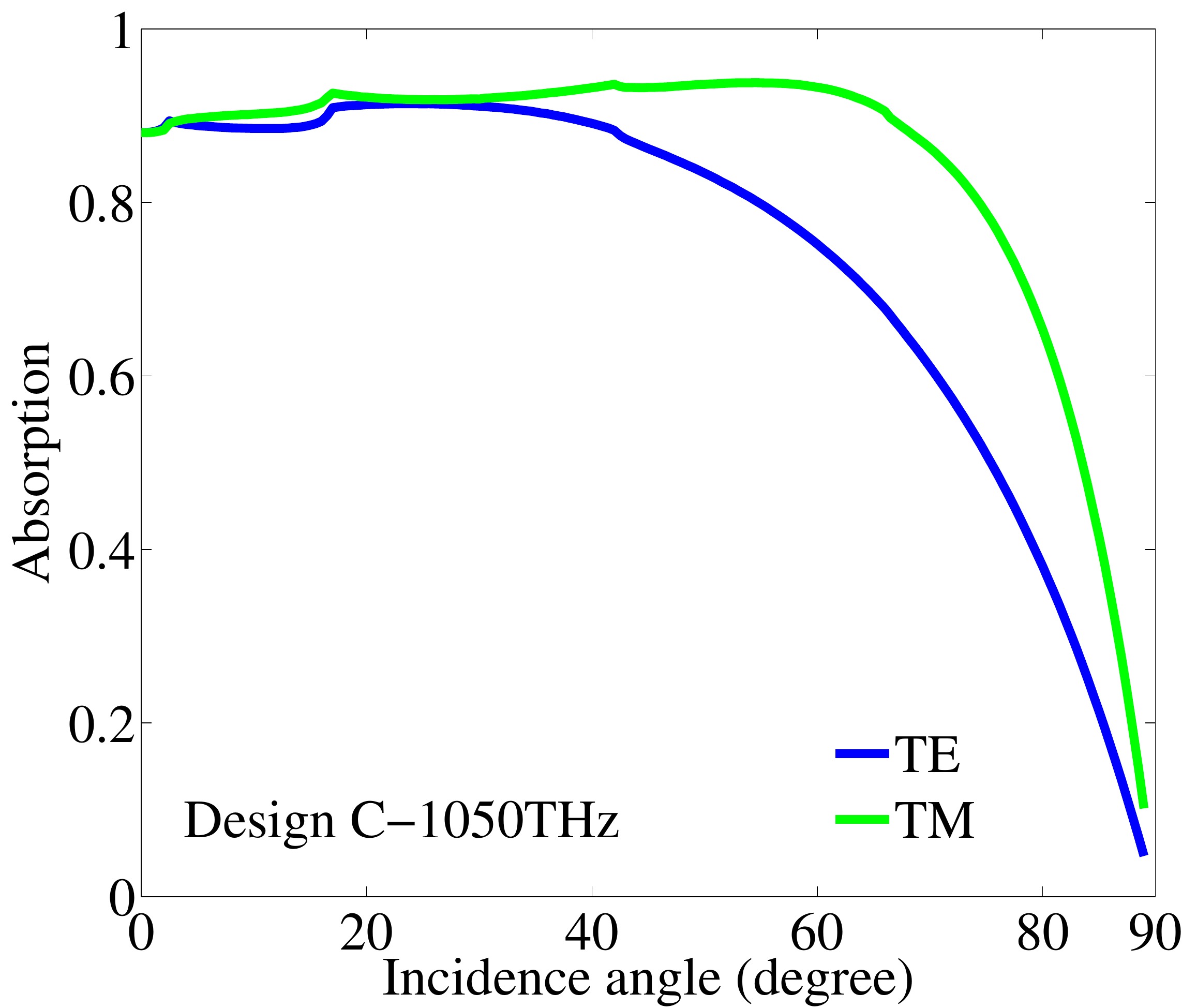} \label{AOS-AD-1050THz}}
\subfigure[]{\includegraphics[width=0.23\textwidth]{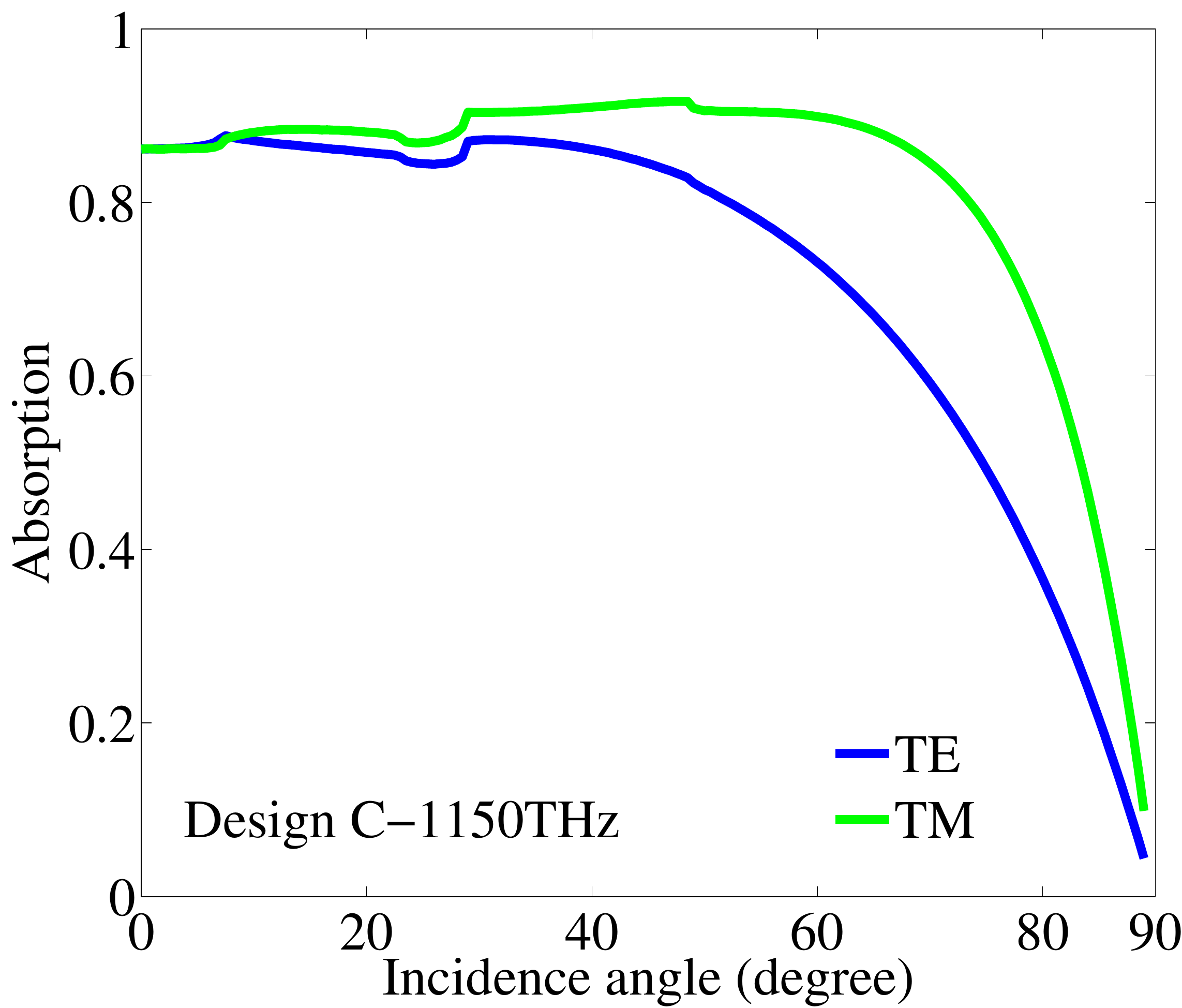} \label{AOS-AD-1150THz}}\\
\subfigure[]{\includegraphics[width=0.27\textwidth]{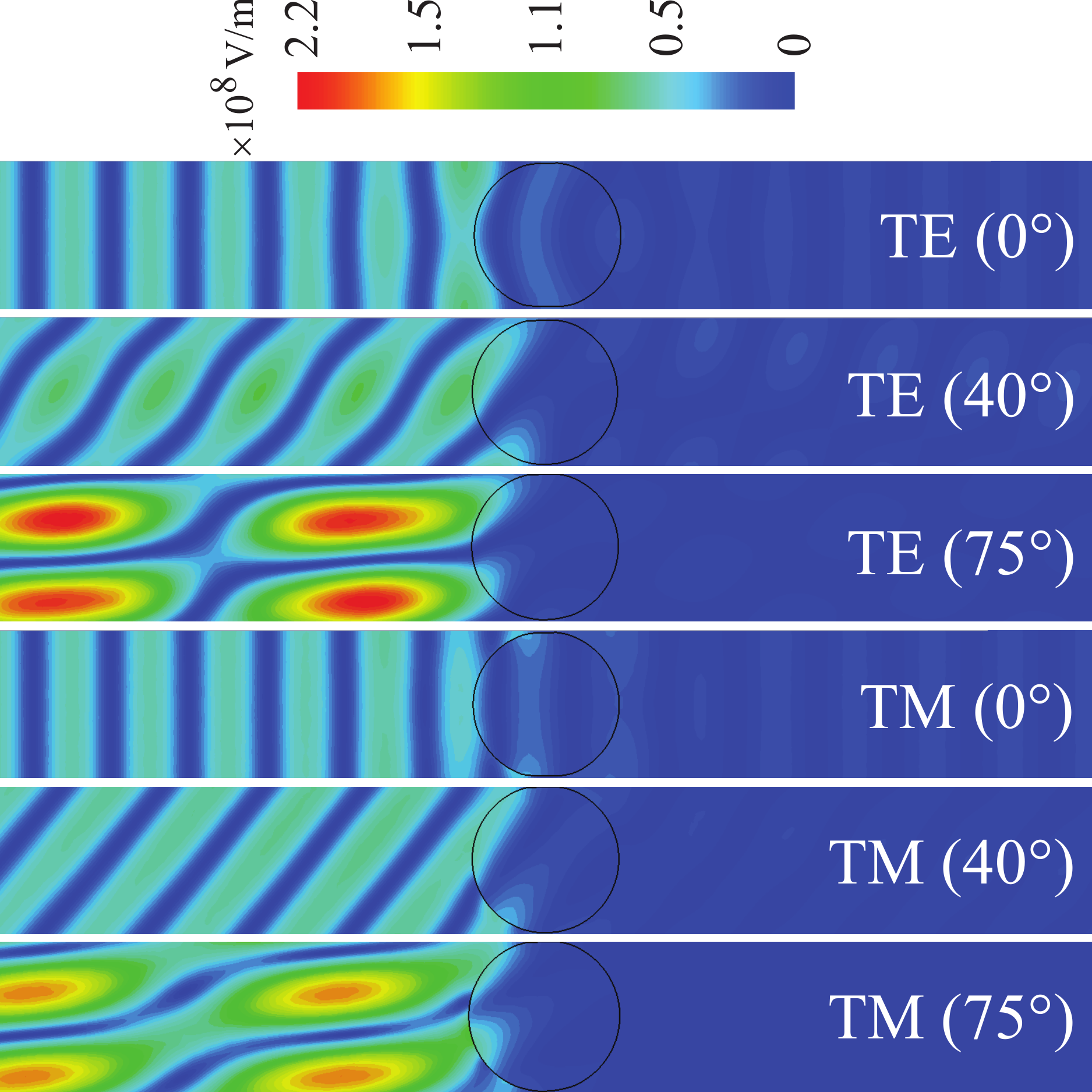} \label{AOS-W-950}}
\caption{Design C: (a) Geometry of the proposed design. (b) Absorption as a function of the frequency. (c) Normalized effective electric and magnetic polarizabilities. (d) and (e) Electric current distributions with $\pi/2$ phase shift for 1000 THz. Absorption as a function of the incidence angle for (f) 950 THz, (g) 1050 THz, and (h) 1150 THz. (i) Electric field distributions (time-harmonic regime, field strength at a fixed moment of time is plotted) for TE (in the E-plane) and TM (in the H-plane) for 950 THz.}
\label{Design-C}
\end{figure}

{\bf Design C.} In the previous design one can notice that $t_{\rm n-Si}\ll r_{\rm Ag}$. This brings up an interesting question: Would it be possible to design a broadband absorber being formed by a single array of silver spheres without any shells? Considering the required conditions for the unit cells of an absorber in Eq.~(5), this is tantamount to asking if it would be possible to bring the electric and magnetic responses of a single silver sphere to balance. Next we demonstrate that arranging bare silver nanospheres of moderate optical sizes in regular arrays it is possible to find such dimensions so that the artificial magnetic response be properly balanced with the electric one, ensuring total absorption. We design an array of silver nanospheres, shown in Fig.~\ref{AOS-Schem}, with $r_{\rm Ag}=148$ nm and $d=300$ nm. Absorption spectrum for this design is shown in Fig.~\ref{AOS-A}, and it exhibits a broad absorption band. As expected, the normalized effective electric and magnetic polarizabilities for silver spheres in the array are balanced and satisfy the required condition defined by Eq.~(3) [see Fig.~\ref{EPolar-C}]. Electric current distributions at 1000 THz shown in Figs.~\ref{AOS-D-E-1000THz} and \ref{AOS-D-M-1000THz} are evidences for the presence of both electric and magnetic responses. It can be seen that in this case the magnetic moment currents are strongly confined in between the neighboring nanoparticles. Absorption angular stability at  different frequencies inside the absorption band, at 950 THz, 1050 THz and 1150 THz, is demonstrated in Figs.~\ref{AOS-AD-950THz}, \ref{AOS-AD-1050THz}, and \ref{AOS-AD-1150THz}, respectively. This design also shows a very good angular stability, as it can also be seen from the electric field distribution in Fig.~\ref{AOS-W-950}.

\begin{figure}[h!]
\centering
\subfigure[]{\includegraphics[width=0.3\textwidth]{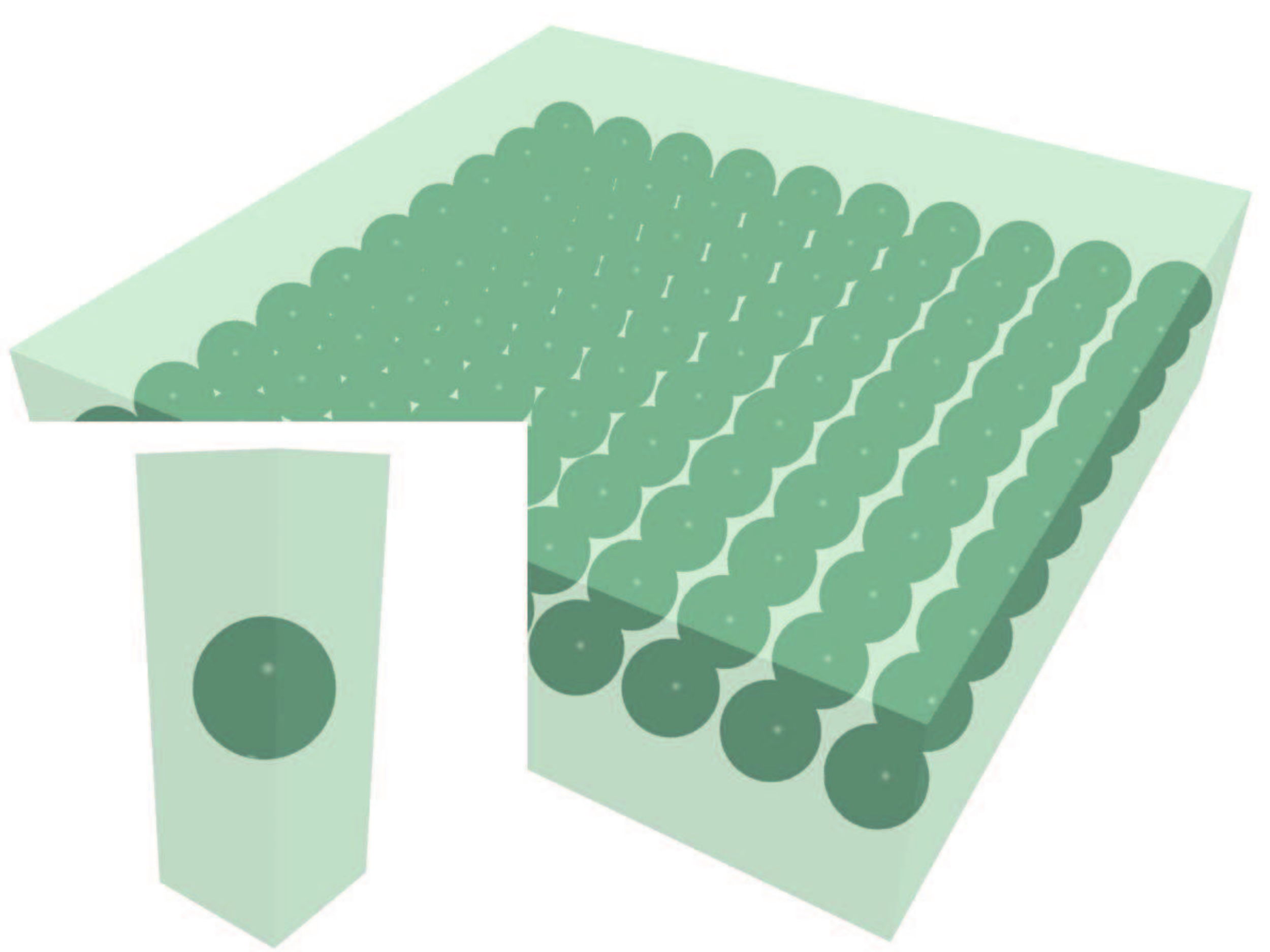} \label{SS-Schem}}
\subfigure[]{\includegraphics[width=0.28\textwidth]{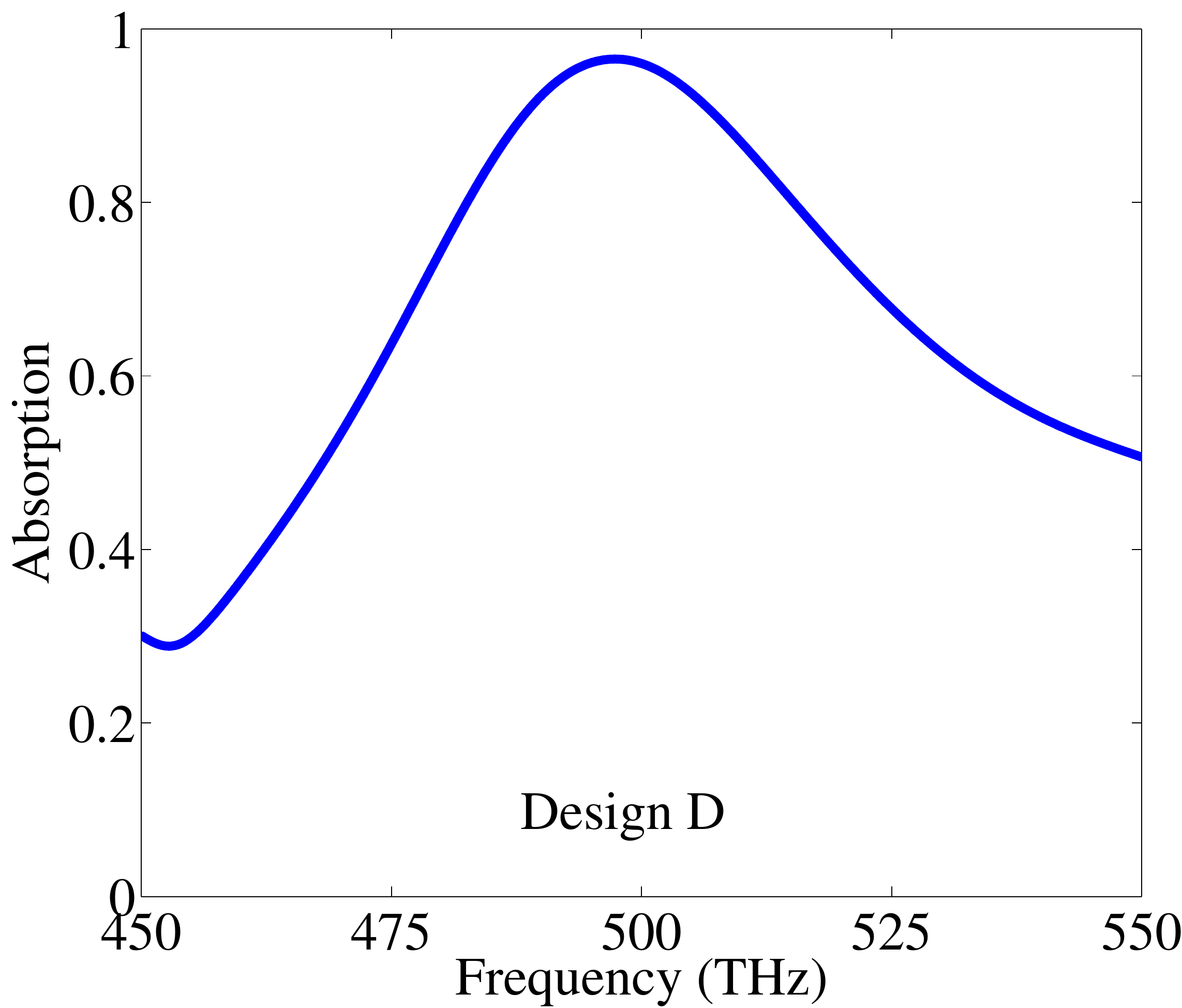} \label{SS-A}}
\subfigure[]{\includegraphics[width=0.33\textwidth]{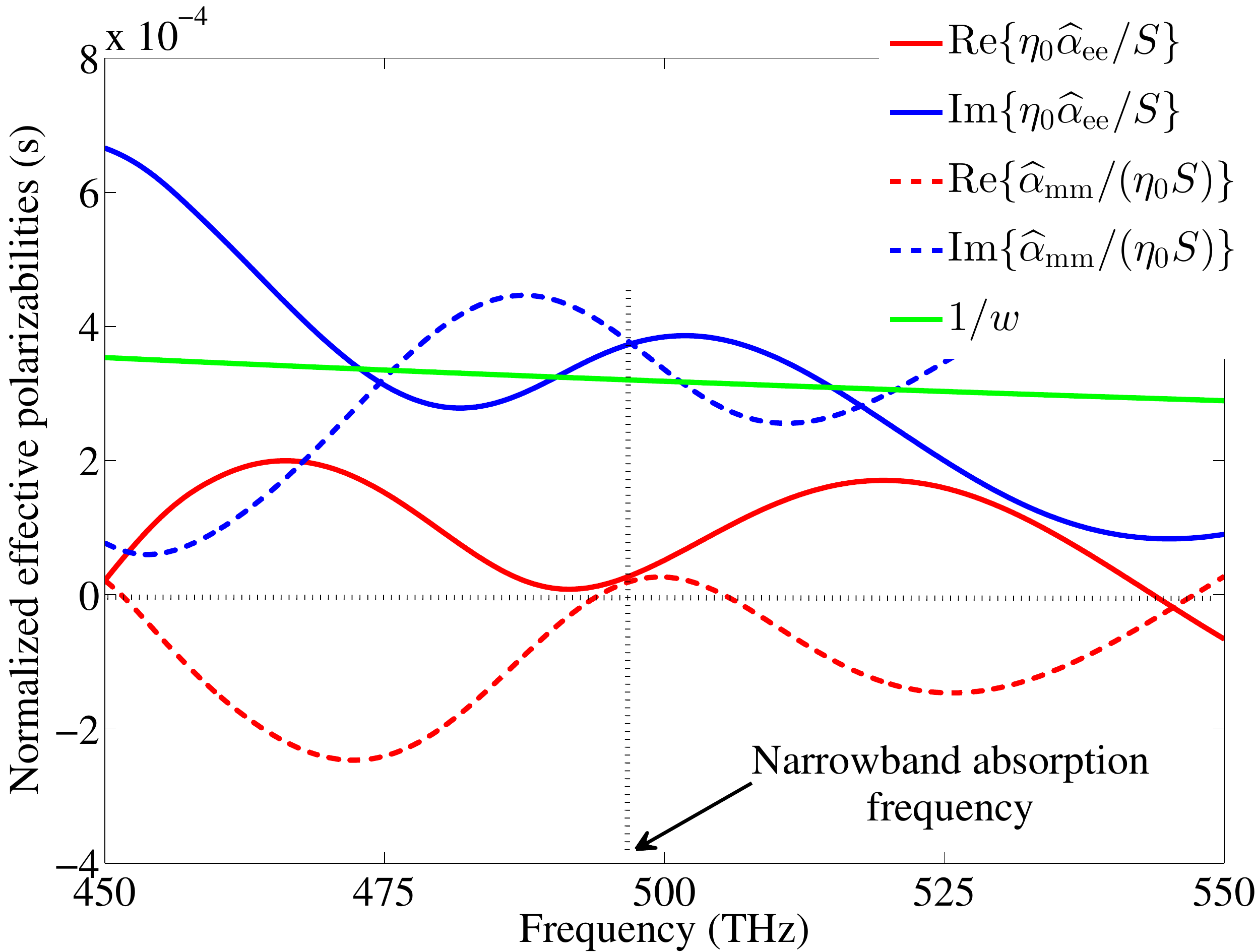} \label{EPolar-D}}
\subfigure[]{\includegraphics[width=0.18\textwidth]{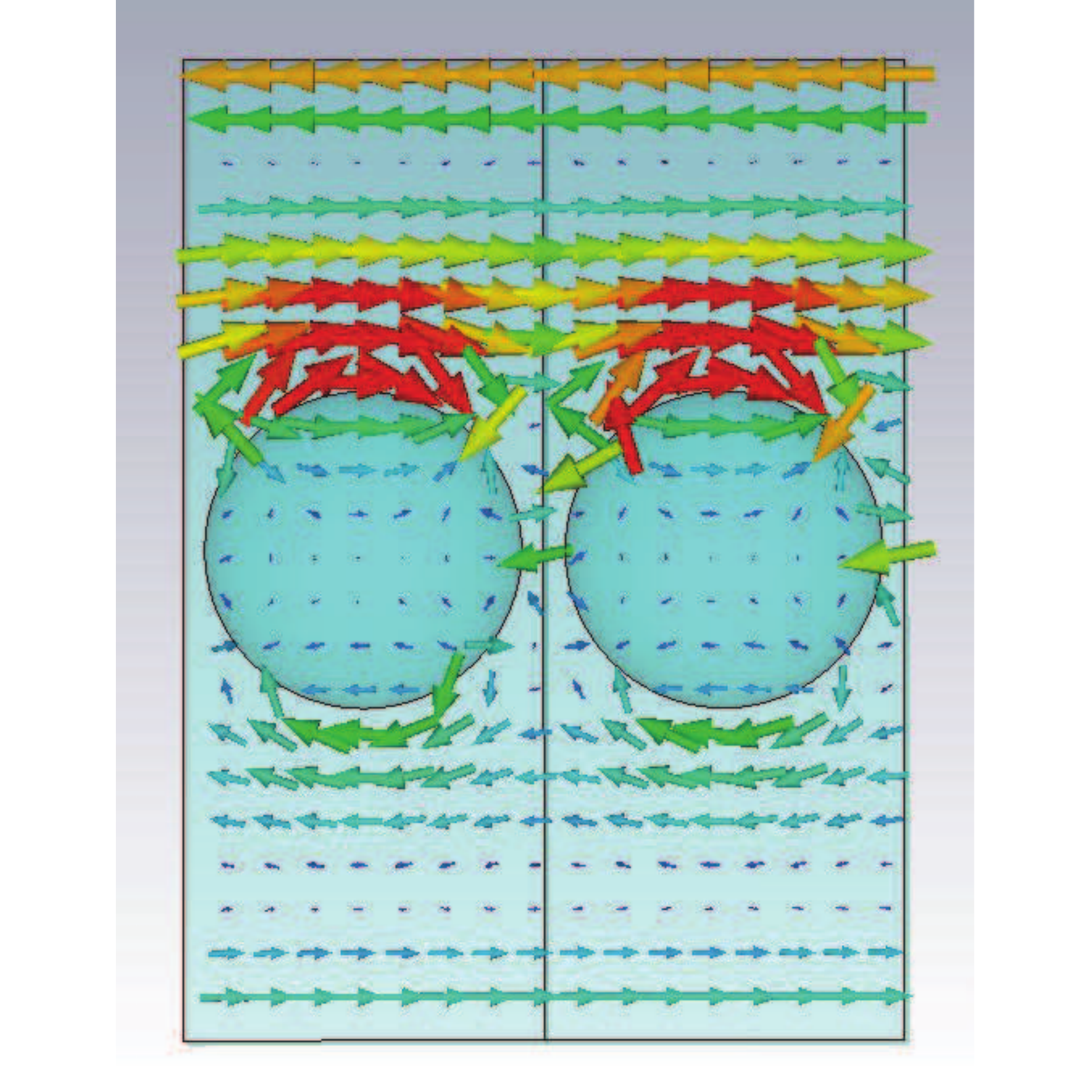} \label{SS-D-E}}
\subfigure[]{\includegraphics[width=0.18\textwidth]{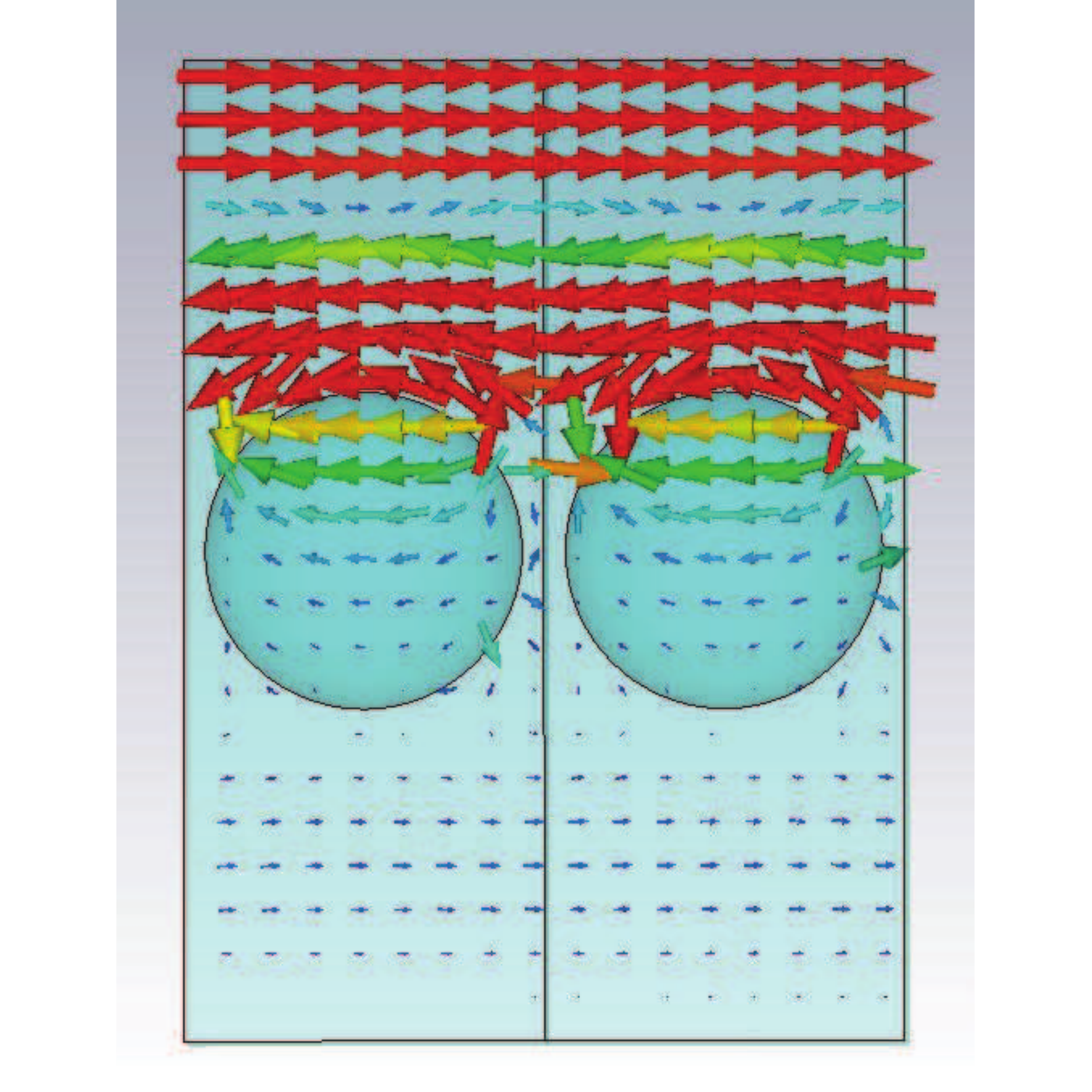} \label{SS-D-M}}
\subfigure[]{\includegraphics[width=0.28\textwidth]{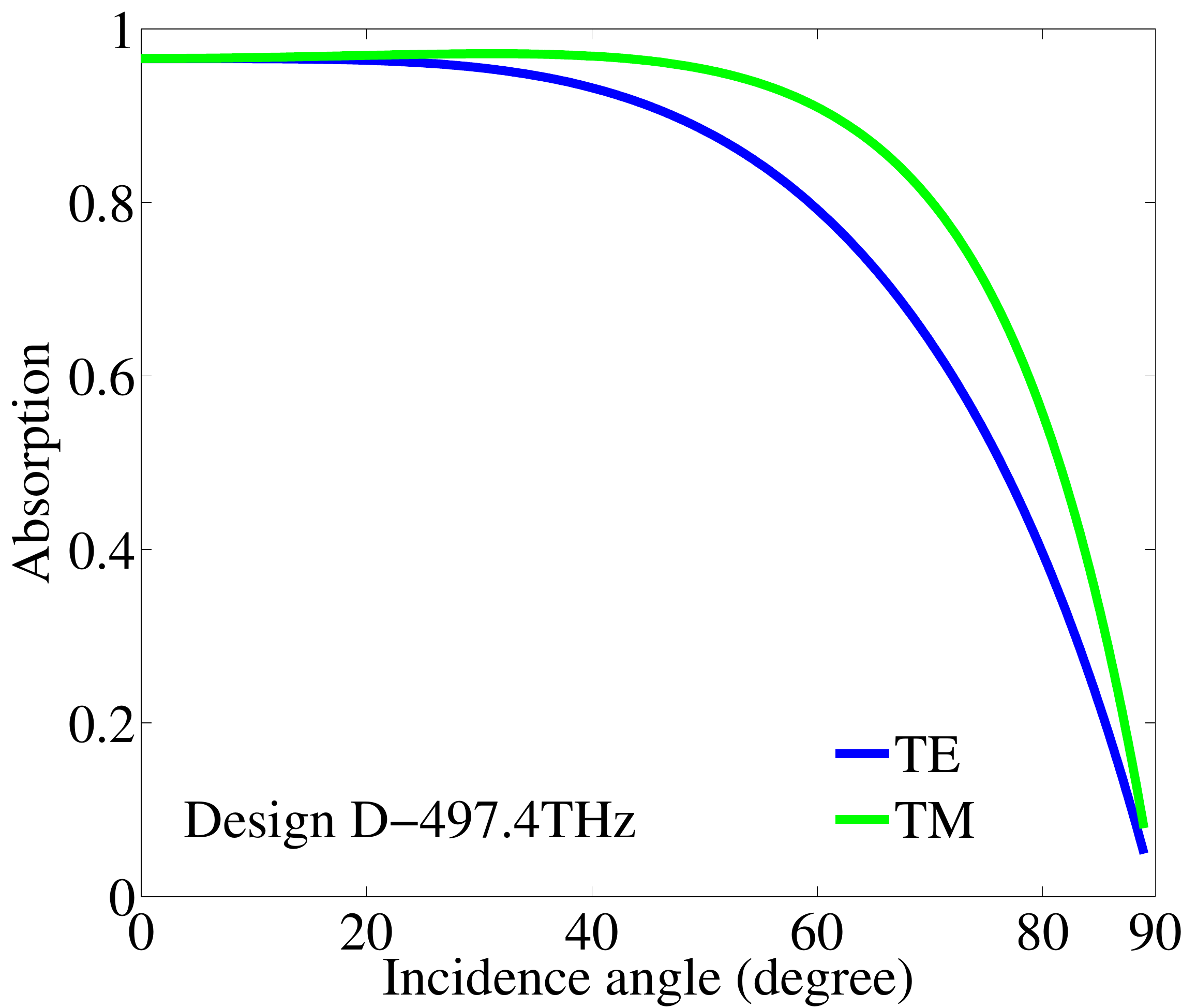} \label{SS-AD}}
\subfigure[]{\includegraphics[width=0.27\textwidth]{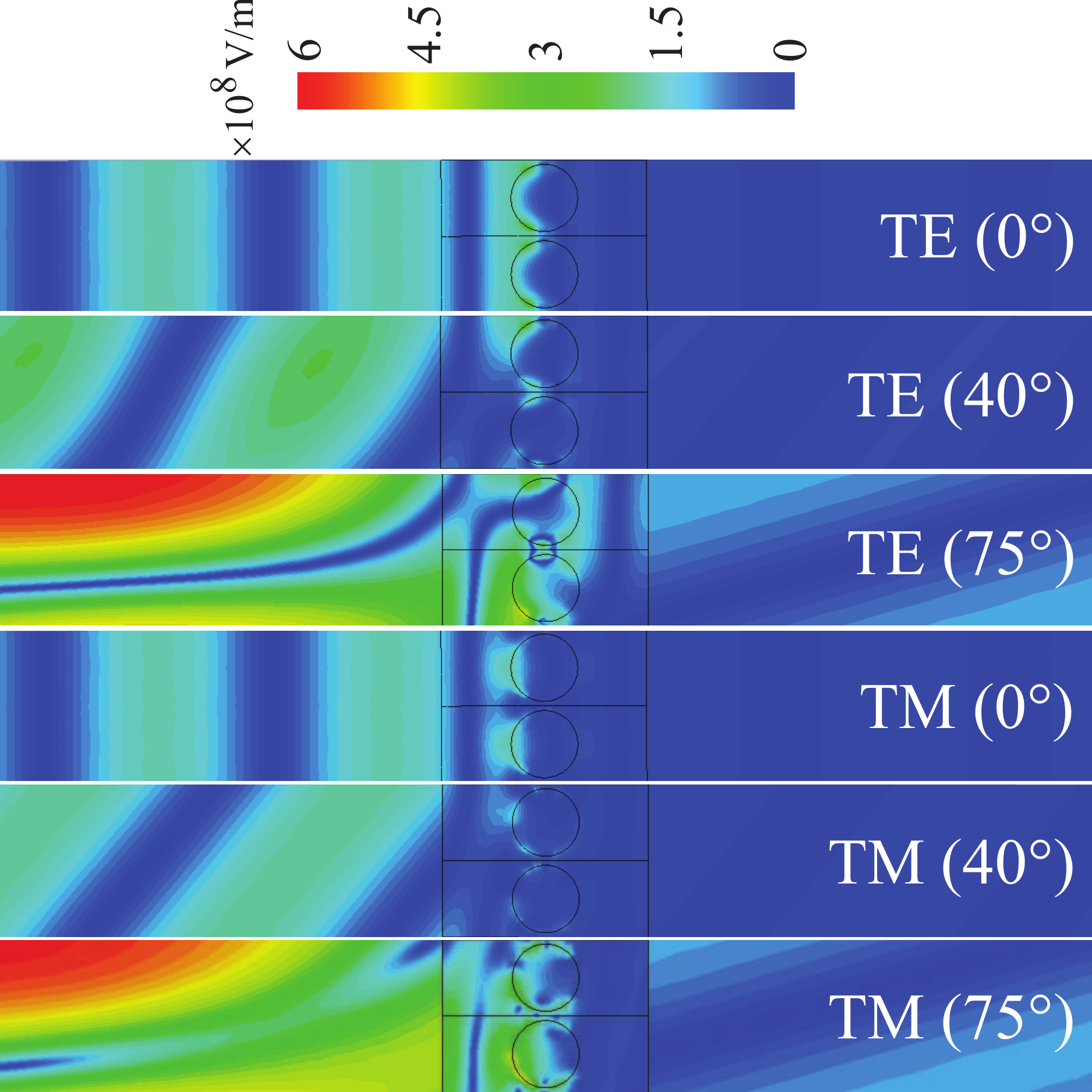} \label{SS-W}}
\caption{Design D: (a) Geometry of the proposed design. (b) Absorption as a function of the frequency. (c) Normalized effective electric and magnetic polarizabilities. (d) and (e) Electric current distributions with $\pi/2$ phase shift for 497.4 THz. (f) Absorption as a function of the incidence angle for 497.4 THz. (g) Electric field distributions (time-harmonic regime, field strength at a fixed moment of time is plotted) for TE (in the E-plane) and TM (in the H-plane) for 497.4 THz.}
\label{Design-D}
\end{figure}

\begin{figure}[h!]
\centering
\includegraphics[width=0.45\textwidth]{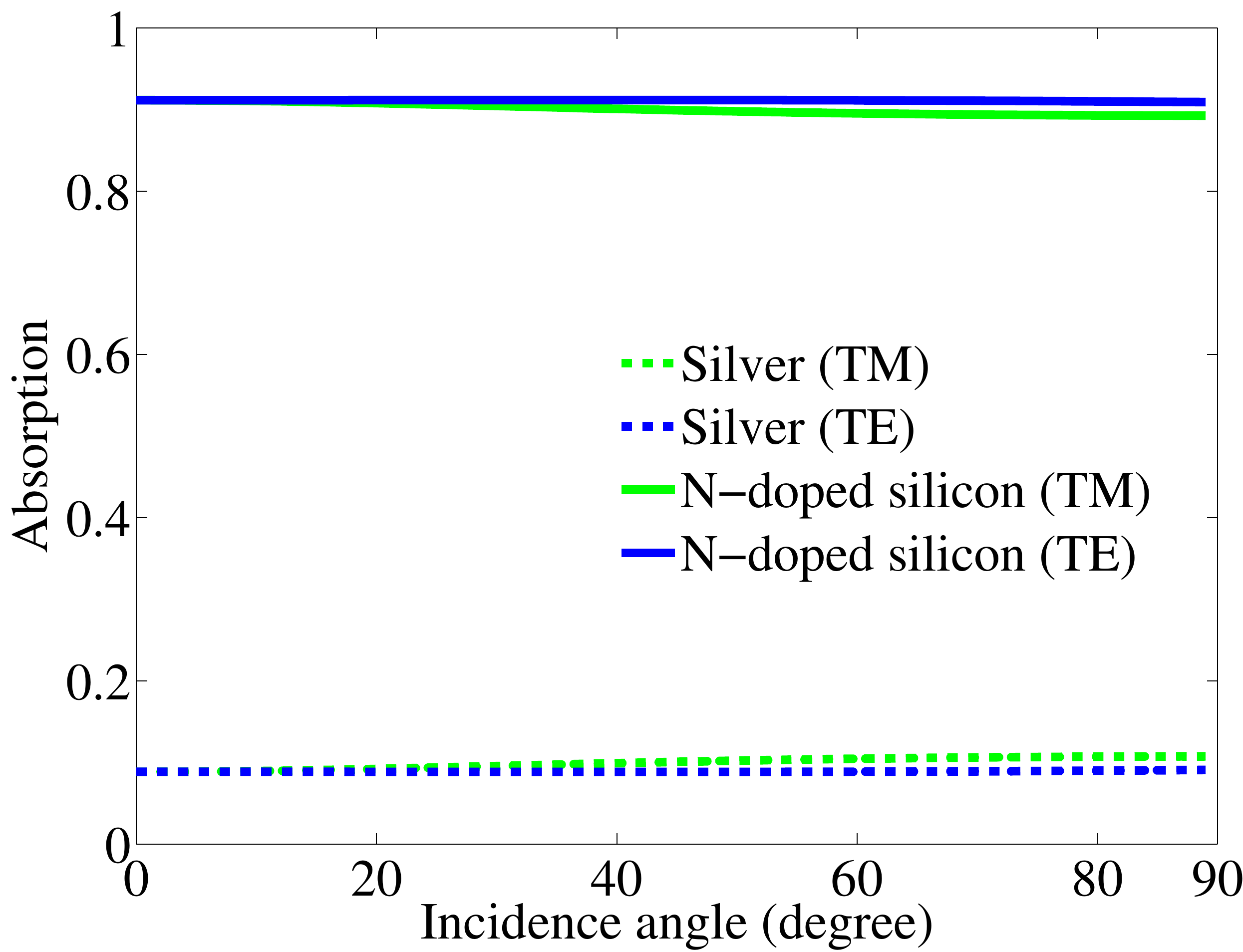} 
\caption{Comparison of  dissipation levels in the silver core and the n-doped silicon shell (silver spheres in a semiconductor slab).}
\label{Ave-SS}
\end{figure}

{\bf Design D.} Figure~\ref{SS-Schem} shows a practically interesting design inspired by the properties of the core-shell particle arrays. In this design, instead of covering silver spheres with n-doped silicon shells, an array of silver spheres is positioned inside an amorphous n-doped silicon slab. Here, the spheres radius, the thickness of the semiconductor slab, and the array period are chosen to be $r_{\rm Ag}=44$ nm, $t_{\rm n-Si}=272$ nm, and $d=100$ nm, respectively. More than 96\% absorption is achieved at $497.4$ THz [Fig.~\ref{SS-A}]. As it can be seen from Fig.~\ref{EPolar-D}, this absorption is achieved due to presence of balanced and furthermore sufficiently strong [satisfying Eq.~(3)] electric and magnetic responses in the layer. The presence of electric and artificial magnetic moments in the structure can also be seen from Figs.~\ref{SS-D-E} and \ref{SS-D-M}. Figure~\ref{SS-AD} shows the angular dependency of absorption at the frequency of the maximum absorption. Because the inclusions and the array period are optically small, there are no higher order modes propagating for any incidence angle. Moreover, the inclusions are spherically symmetric and respond as electric and magnetic dipoles. That is why the structure shows reasonably high angular stability for both TE and TM polarizations. The electric field distributions for the maximum absorption frequency are reported in Fig.~\ref{SS-W}. Figure~\ref{Ave-SS} compares the power dissipated in the semiconductor and metal fractions for different polarizations as a function of the incidence angle. It can be seen that about 91\% of absorbed power is dissipated inside the n-doped silicon. This result is quite important from the practical point of view, especially for energy harvesting applications.

In should be noted that in all these designs the induced current distributions indicate that the amplitudes of higher-order multipoles are significant. However, for the normal incidence the higher-order multipoles do not contribute to the reflected and transmitted plane waves. Their presence affects the performance only through modification of the values of the effective electric and magnetic polarizabilities. For
oblique incidence the higher-order modes are of more significance and somewhat affect the operation
of the absorbers.

Now let us discuss the manufacturing feasibility of the suggested absorbing monolayers of nanoparticles.
The most difficult task is fabrication of core-shell particles with a metal core and a semiconductor shell Figs.~\ref{N-Schem} and~\ref{B-Schem}. In spite of significant interest for numerous possible applications, preparation of nanoparticles with silver core and silicon shell has been demonstrated only by a few scientific groups (see e.g. \cite{Buryatia}), unlike nanospheres with silicon core and silver shell \cite{commonplace} or silver core and silica shell \cite{Quinsaat} whose technologies are referred as ``conventional'' in the modern nanochemistry \cite{Nanochemistry}. The novel evaporation-condensation technology suggested in \cite{Buryatia} has been developed and approaches more widespread utilization (see patent \cite{patent}). An alternative technology of producing such particles is purely chemical \cite{Misha}. It requires a colloid of silver cores covered with silica shells \cite{Quinsaat} which are transformed into silicon shells. The reaction transforming silica nanospheres into silicon ones was performed in \cite{Misha}. High chemical stability of silver to silicon and silicates means that the presence of the silver core inside silica will not compromise  this mechanism.

Once nanoparticles are obtained in a colloid, they can be assembled into a densely packed monolayer on a transparent substrate using either chemical or physical technologies. Namely, it can be a top-down technique \cite{Misha1}, which is very time-consuming but guarantees a very high regularity of the array. Alternatively, it can be a standard self-assembly \cite{Amor}. This approach results in dense packaging of particles \cite{Amor}, whereas in Figs.~\ref{N-Schem} and~\ref{B-Schem} the core-shell particles are separated with predefined gaps. However, a nanoparticle with either metal or semiconductor shell may be additionally encapsulated by silica or polystyrene \cite{Quinsaat}. Then the dense package of these double-shell particles will determine the separation of the silver-silicon nanoparticles, whereas the role of the second shell will be only to ensure the needed particle separation. The same encapsulation can be used to prepare a monolayer of simple silver nanoparticles with a predefined period as in Fig.~\ref{AOS-Schem}. Low-index shells weakly disturb the operation of the array \cite{Quinsaat} and most probably will only slightly shift the operation frequency. In any case the optical contrast of low-index shells can be easily compensated by pouring glycerol on top of the array. This liquid has the same refractive index as that of silica or polystyrene and its presence will also remove the optical contrast of the silica substrate.

To fabricate the structure shown in Fig.~\ref{SS-Schem} one may apply the technology described in \cite{Misha2}, where the self-assembly of plasmonic nanoparticles on a silicon substrate is combined with their passivation by organic molecules. The thickness of this organic shell may be controlled within the interval 1-5 nm \cite{Misha2}. This interval of thicknesses corresponds to the separation of nanoparticles implied in Fig.~\ref{AOS-Schem}. After preparing the self-assembled monolayer on a silicon substrate it may be covered by amorphous silicon using one of two standard silicon deposition technologies: sputtering or PECVD \cite{Ohring}. In the first case the organic nanoshells will be preserved whereas all the voids will be filled with amorphous silicon and the result will be close to that shown in Fig.~\ref{SS-Schem} with the only difference: silver spheres will have negligibly thin shells. In the second case the organic ligands will be thermally destroyed and enter the host amorphous silicon as a low-density random mixture. In both cases the fraction of organic molecules in the metamaterial layer will be very small, and its impact on the optical properties of the structure will be, most probable, negligible.

\section{Conclusion}
In summary, we have introduced single-layer arrays of spherical core-shell nanoparticles as absorbers of visible light. The proposed structures have no continuous metallic reflector and effectively absorb light hitting either or both sides of the metasurface. We have demonstrated a number of designs of resonant and broadband absorbers, including structures which do not require any shells over metal spheres, and have shown that in the metal-semiconductor designs most of the power is absorbed in semiconductor and not lost in metal.  We believe that such unique features of the  proposed single-layer absorbing arrays as ultra-wideband, angularly stable and symmetric absorption can open various new application perspectives, especially in energy harvesting devices.


\begin{thebibliography}{00}

\bibitem{Huygens1}
F.~Monticone, N.~M.~Estakhri, and A.~Al\`u, Full control of nanoscale optical transmission with a composite metascreen, Phys. Rev. Lett. {\bf 110}, 203903 (2013).

\bibitem{Huygens2}
C.~Pfeiffer and A.~Grbic, Metamaterial Huygens' surfaces: tailoring wave fronts with reflectionless sheets, Phys. Rev. Lett. {\bf 110}, 197401 (2013).

\bibitem{Transparency}
Y.~Ra'di, V.~S.~Asadchy, S.~A.~Tretyakov, One-way transparent sheets, Phys. Rev. B {\bf 89}, 075109 (2014).

\bibitem{Metamirrors}
Y.~Ra'di, V.~S.~Asadchy, S.~A.~Tretyakov, Tailoring reflections from thin composite metamirrors, IEEE Trans. Antennas Propag. {\bf 62}, 3749 (2014).

\bibitem{Padilla}
C.~M.~Watts, X.~Liu, and W.~J.~Padilla, Metamaterial electromagnetic wave absorbers, Adv. Mater. {\bf 24}, OP98 (2012).

\bibitem{Ramakrishna}
G.~Dayal and S.~A.~Ramakrishna, Design of highly absorbing metamaterials for infrared frequencies, Opt. Exp. {\bf 20}, 17503 (2012).

\bibitem{Atwater}
K.~Aydin, V.~E.~Ferry, R.~M.~Briggs, and H.~A.~Atwater, Broadband polarization-independent resonant light absorption using ultrathin plasmonic super absorbers, Nat. Comm. {\bf 2}, 517 (2011).

\bibitem{Yu}
Y.~Yu, V.~E.~Ferry, A.~P.~Alivisatos, and L.~Cao, Dielectric coreâshell optical antennas for strong solar absorption enhancement, Nano Lett. {\bf 12}, 3674 (2012).

\bibitem{Polman}
H.~A.~Atwater and A.~Polman, Plasmonics for improved photovoltaic devices, Nat. Mater. {\bf 9}, 205 (2010).

\bibitem{TPV}
M.~Laroche, R.~Carminati, and J.~-J.~Greffet, Near-field thermophotovoltaic energy conversion, J. Appl. Phys. {\bf 100}, 063704 (2006).

\bibitem{He}
Y.~Q.~Ye, Y.~Jin, and S.~He, Omnidirectional, polarization-insensitive and broadband thin absorber in the terahertz regime, J. Opt. Soc. Am. B  {\bf 27}, 498 (2010).

\bibitem{Fang}
Y.~Cui, K.~H.~Fung, J.~Xu, H.~Ma, Y.~Jin, S.~He, and N.~X.~Fang, Ultrabroadband light absorption by a sawtooth anisotropic metamaterial slab, Nano Lett. {\bf 12}, 1443 (2012).

\bibitem{BackedCS1}
C.-C.~Lee and D.-H.~Chen, Ag nanoshell-induced dual-frequency electromagnetic wave absorption of Ni nanoparticles, Appl. Phys. Lett. {\bf 90}, 193102 (2007).

\bibitem{BackedCS2}
C.~H\"agglund and S.~P.~Apell, Plasmonic near-field absorbers for ultrathin solar cells, J. Phys. Chem. Lett. {\bf 3}, 1275 (2012).

\bibitem{BackedCS3}
J.~Dai, F.~Ye, Y.~Chen, M.~Muhammed, M.~Qiu, and M.~Yan, Light absorber based on nano-spheres on a substrate reflector, Opt. Exp. {\bf 21}, 6697 (2013).

\bibitem{absorption}
Y.~Ra'di, V.~S.~Asadchy, and S.~A.~Tretyakov, Total absorption of electromagnetic waves in ultimately thin layers, IEEE Trans. Antennas Propag. {\bf 61}, 4606 (2013).


\bibitem{coherent0}
W. Wan, Y. Chong, L. Ge, H. Noh, A. D. Stone, and H. Cao,
Time-reversed lasing and interferometric control of absorption,
Science {\bf 331}, 889 (2011).

\bibitem{coherent1}
M.~Kang, F.~Liu, T.-F.~Li, Q.-H.~Guo, J.~Li, and J.~Chen, Polarization-independent coherent perfect absorption by a dipole-like metasurface, Opt. Lett. {\bf 38}, 3086 (2013).

\bibitem{coherent2}
J.~Zhang, K.~F. MacDonald, and N.~I. Zheludev, Controlling light-with-light without nonlinearity, Light: Science \& Applications {\bf 1}, e18 (2012).

\bibitem{Viicha}
V.~S.~Asadchy, I.~A.~Faniayeu, Y.~Ra'di, S.~A.~Khakhomov, I.~V.~Semchenko, and S.~A.~Tretyakov, Broadband reflectionless metasheets: Frequency-selective transmission and perfect absorption, arXiv:1502.06916.

\bibitem{Costas_cylinders}
C.~A.~Valagiannopoulos and S.~A.~Tretyakov, Symmetric absorbers realized as gratings of PEC cylinders covered by ordinary dielectrics, IEEE Trans. Antennas and Propag. {\bf 62}, 5089 (2014).

\bibitem{Samofalov}
I.~V.~Semchenko, S.~A.~Khakhomov, and A.~L.~Samofalov, Helices of optimal shape for nonreflecting covering, Eur. Phys. J. Appl. Phys. {\bf 49}, 33002 (2010).


\bibitem{Teemu}
T.~Niemi, A.~O.~Karilainen, and S.~A.~Tretyakov, Synthesis of polarization transformers, IEEE Trans. Antennas Propag. {\bf 61}, 3102 (2013).


\bibitem{miss}
R.~Paniagua-Dominguez, F. Lopez-Tejeira, R. Marques, and J.~A.~Sanchez-Gil, Metallo-dielectric coreâshell nanospheres as building blocks for optical three-dimensional isotropic negative-index metamaterials, New J. Phys. {\bf 13},  123017 (2011).

\bibitem{Morits}
D.~Morits and C.~R.~Simovski, Isotropic negative refractive index at near infrared, J. Optics {\bf 14}, 125102 (2012).

\bibitem{Doyle}
W.~T.~Doyle, Optical properties of a suspension of metal spheres, Phys. Rev. B {\bf 39}, 9852 (1989).

\bibitem{Johnson}
P.~B.~Johnson and R.~W.~Christy, Optical Constants of the Noble Metals, Phys. Rev. B {\bf 6}, 4370 (1972).

\bibitem{Ulrich}
U.~Stutenb\"aumer, B.~Mesfin, and S.~Beneberu, Determination of the optical constants and dielectric functions of thin film a-Si: H solar cell layers, Solar Energy Materials and Solar Cells {\bf 57}, 49 (1999).

\bibitem{CST}
CST Studio Suite, 2013: http://www.cst.com.

\bibitem{Jokerst}
X.~Liu, T.~Tyler, T.~Starr, A.~F.~Starr, N.~M.~Jokerst, and W.~J.~Padilla, Taming the blackbody with infrared metamaterials as selective thermal emitters, Phys. Rev. Lett. {\bf 107}, 045901 (2011).

\bibitem{Mayer}
Z.~H.~Jiang, S.~Yun, F.~Toor, D.~H.~Werner, and T.~S.~Mayer, Conformal dual-band near-perfectly absorbing mid-infrared metamaterial coating, ACS Nano {\bf 5}, 4641 (2011).

\bibitem{Buryatia}
A.~V.~Nomoev and S.~P.~Bardakhanov, Synthesis and structure of AgSi nanoparticles obtained by the electron-beam evaporation/condensation method,  Technical Physics Letters, \textbf{38},  375 (2012).

\bibitem{commonplace}
J.~Zhang, Y.~Fu, and J.~R.~Lakowicz, Fluorescent metal nanoshells: lifetime-tunable molecular probes in fluorescent
cell imaging, Journal of Physical Chemistry C, \textbf{115}, 7255 (2011).

\bibitem{Quinsaat}
J.~E.~Q.~Quinsaat, F.~A.~Nuesch, H.~Hofmann, and D.~M.~Opris, Dielectric properties of silver nanoparticles coated with silica shells of different thicknesses, RSC Advances \textbf{3}, 6964 (2013).

\bibitem{Nanochemistry}
A.~K.~Haghi and G.~Zaikov, \emph{Modern Nanochemistry}, Nova Science, 2012.

\bibitem{patent}
D.~S.~Vidya, C.~Cassidy, and M.~I.~Sowwan, Metal induced nanocrystallization of amorphous semiconductor quantum dots, US Patent WO2014141662 A1 (Sept. 18, 2014).

\bibitem{Misha}
W.~Wang, Z.~Favors, R.~Ionescu, R.~Ye, H.~H.~Bay, M.~Ozkan, and C.~S.~Ozkan, Monodisperse porous silicon spheres as anode materials for lithium ion batteries, Sci. Rep. \textbf{5}, 8781 (2015).

\bibitem{Misha1}
Y.~Bao , Z.~Yan , and N.~F.~Scherer, Optical printing of electrodynamically coupled metallic nanoparticle arrays, J. Phys. Chem. C {\bf 118}, 19315 (2014).

\bibitem{Amor}
C.~Rockstuhl and T.~Scharf (Eds.), \emph{Amorphous Nanophotonics,} Springer, 2013.

\bibitem{Misha2}
V.~Santhanam, J.~Liu, R.~Agarwal, and R.~P.~Andres, Self-assembly of uniform monolayer arrays of nanoparticles, Langmuir {\bf 19}, 7881 (2003).

\bibitem{Ohring}
M.~Ohring, \emph{Materials Science of Thin Films}, 2 ed., Academic Press (2002).


\end{thebibliography}
\end{document}